\documentclass[journal=jpcafh,manuscript=article,hyperref=true]{achemso}

\setkeys{acs}{doi=true,keywords=true}

\hypersetup{hidelinks}

\usepackage{amsmath}
\usepackage{amssymb}
\usepackage{booktabs}
\usepackage{enumitem}
\usepackage[section]{placeins}

\SectionNumbersOn

\newcommand{\phig}{\varphi}
\newcommand{\chistruct}{\chi_{\mathrm{struct}}}
\newcommand{\kappaRS}{\kappa_{\mathrm{RS}}}
\newcommand{\Jchem}{J_{\mathrm{chem}}}
\newcommand{\Kmol}{K_{\mathrm{mol}}}
\newcommand{\chistar}{\chi^{\star}}

\author{Megan Simons}
\email{msimons@recognitionphysics.org}
\phone{+1-512-766-8363}
\affiliation[Recognition Physics Institute]
  {Recognition Physics Institute, 12710 Research Blvd \#210,
   Austin, Texas 78759, USA}
\author{Jonathan Washburn}
\affiliation[Recognition Physics Institute]
  {Recognition Physics Institute, 12710 Research Blvd \#210,
   Austin, Texas 78759, USA}
\author{Elshad Allahyarov}
\affiliation[Recognition Physics Institute]
  {Recognition Physics Institute, 12710 Research Blvd \#210,
   Austin, Texas 78759, USA}
\alsoaffiliation[Case Western Reserve University]
  {Department of Physics, Case Western Reserve University, 10900 Euclid Avenue, Cleveland, Ohio 44106, USA}
\alsoaffiliation[Heinrich-Heine-Universit\"at D\"usseldorf]
  {Institut f\"ur Theoretische Physik II, Heinrich-Heine-Universit\"at
   D\"usseldorf, Universit\"atsstra\ss{}e 1, 40225 D\"usseldorf, Germany}
\alsoaffiliation[Joint Institute for High Temperatures]
  {Theoretical Department, Joint Institute for High Temperatures, Russian
   Academy of Sciences, Izhorskaya 13 Bldg 2, Moscow 125412, Russia}

\title[Noble-Gas Coordinate in Equalization]
  {Molecular Transferability of a Noble-Gas Coordinate for Electronegativity Equalization}


\newcommand{\tabnotes}[1]{%
  \par\vspace{2pt}%
  \parbox{0.97\textwidth}{\footnotesize\raggedright #1}%
}

\begin{document}

\singlespacing

\begin{abstract}
Charge-equilibration models predict how electrons redistribute over a molecule more cheaply than quantum chemistry, and each begins from a table of atomic electronegativities and hardnesses. A recent table replaces ionization energies and electron affinities with one geometric quantity: each main-group atom's fractional distance to the noble gas closing its row. Holding geometries, electrostatic damping and the molecule list fixed, we compare it with spectroscopic values, with its own kernel refitted, and with a quadratic control, over a primary set of fifty-two molecules and ions at the B3LYP/def2-TZVP level. Two structural consequences follow from the table construction rather than the equalization solver. Atoms at equal fractional distance receive identical electronegativities, so the dipole moments of chlorine monofluoride, iodine monobromide and sulfur dioxide vanish exactly, and the measured 0.72 debye HF/HCl gap is lost. Because the first period is excluded, hydrogen retains its spectroscopic electronegativity above every geometric value, reversing O--H, N--H and hydrogen-halide polarity. Fitted alike, geometric and quadratic kernels agree to 0.0022 electron per atom, so predictions are set by the fitted energy scales, not the kernel shape. Resolving the electronegativity scale period by period restores correct polarity in all five polyatomic heavy-atom--hydrogen tests.
\end{abstract}

\maketitle

\begin{tocentry}
\includegraphics[width=\linewidth]{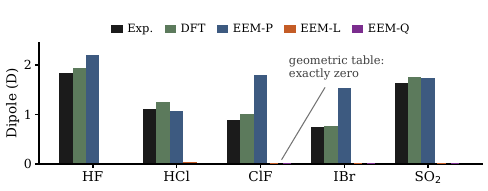}
\end{tocentry}

\section{Introduction}
\label{sec:intro}

Molecular electrostatics is frequently needed faster than a quantum-chemical calculation can supply it, and charge-equilibration models are designed for exactly that situation: they distribute charge across a molecule by demanding that the chemical potentials of its atoms become equal, and they supply the electrostatics of fluctuating-charge force fields \cite{rappe1991qeq,verstraelen2009sqe,verstraelen2013acks2}. Every model of this family starts from a table of atomic electronegativities and hardnesses, and it is the table, not the model, that concerns us here. The electronegativity \(\chi\) and the hardness \(\eta\) are properties of the isolated atom, obtained from its ionization energy and electron affinity through Eqs.~\eqref{eq:mulliken}--\eqref{eq:pearson-eta}. Other atomic scales are available (among them those of Pauling, Allred and Rochow, and Allen \cite{pauling1932nature,allred1958electronegativity,allen1989electronegativity}), but none of them can serve as input here, for a reason peculiar to equalization: Eq.~\eqref{eq:energy} requires \(\chi\) and a matching \(\eta\) expressed in energy units and drawn from a single definition. The Pauling scale is constructed from bond energies and is dimensionless; the Allred--Rochow scale estimates an electrostatic force and is then calibrated onto that dimensionless scale; and Allen configuration energies, although expressed in electronvolts, have no hardness counterpart. Atomic values alone do not determine the partial charges within a molecule. Charge flows from one atom to another until the chemical potentials equalize, and the charges that result, together with the dipole moment they generate, are the quantities against which a table must be judged. Of these two, only the dipole moment is an observable, because a partial charge acquires a value only once a partition of the molecular electron density has been chosen (Sec.~\ref{sec:verdicts}). The question at issue is therefore not whether a set of atomic numbers can be written down, but whether a given table, once inserted into an equalization model, yields charges and dipole moments that distinguish fluorine from chlorine and that preserve the conventional polarity of an O--H bond.

Two strategies dominate current practice. The first keeps the number of adjustable parameters small and accepts whatever accuracy follows: the geometric-mean postulate of Sanderson \cite{sanderson1951electronegativity,sanderson1983electronegativity} and the electronegativity equalization method (EEM) of Mortier and coworkers \cite{mortier1985electronegativity,mortier1986electronegativity} both take atomic \(\chi\) and \(\eta\) as they stand. The second abandons the atomic reading of these parameters and fits effective values to quantum-chemical charges over a training set of molecules, which is how the method is usually applied in practice \cite{bultinck2002eem}. Related quadratic models include charge equilibration and later variants that repair the dissociation limit or the scaling of the polarizability \cite{rappe1991qeq,nistor2006split,mathieu2007split,verstraelen2009sqe,verstraelen2013acks2}. All of these variants share a quadratic dependence of the energy on the atomic charges, and the machinery built on that dependence in the charge-sensitivity literature \cite{nalewajski1996charge} is used in Sec.~\ref{sec:theory}. Parameters obtained by fitting depend strongly on the calibration procedure, and least-squares fitting to charges alone can be ill-conditioned \cite{verstraelen2011params}. The present work follows neither strategy. It inserts a table that was fixed before any molecule was considered, which we call a frozen table, and performs no molecular calibration whatsoever, which is what makes the spectroscopic Mulliken--Pearson table the appropriate untuned baseline for comparison. A frozen table is not a parameter-free table: the geometric construction carries five atomic-scale constants fitted to spectroscopic data. The scope and design limitations of the comparison are collected in Sec.~\ref{sec:scope}. One feature of the equalization model itself has to be dealt with before any table can be assessed. If the electrostatic interaction between two atoms is left as a bare Coulomb term, it grows without bound as they approach and at bonded distances it overwhelms the atomic hardnesses, which reverses the sign of the charges in a molecule as simple as hydrogen fluoride. A damped form is therefore used, following Ohno \cite{ohno1964}, and the same form is used for all four tables, so it cannot account for any difference between them; the demonstration is in Sec.~\ref{sec:eem}. We adopt it unchanged for every table, so that no difference between tables can be attributed to the damping. These are properties of the molecular machinery, and they leave open the question of which atomic table should be inserted into it.

The table examined here comes from a recent construction that locates each main-group atom by a single dimensionless coordinate \(\rho\): the distance from the atom to the noble gas that closes its row, divided by the length of that row (Sec.~\ref{sec:rho}). A convex function of \(\rho\) is then differenced twice, once across a single step away from the closing noble gas and once across the remaining width of the row. The first difference is used in place of an ionization energy and the second in place of an electron affinity, so that their mean becomes an electronegativity and the second difference alone becomes a hardness. Three features of this scheme were already apparent for free atoms \cite{washburn2026noblegas}: two of them limitations, and the third a question of whether the published function is identified by the data at all. Fluorine and chlorine occupy the same fractional position and therefore receive the same electronegativity whatever energy scales are adopted, so the construction cannot express the fact that fluorine is the more electronegative of the two. In addition, the first period is excluded from the construction, so hydrogen is never assigned a coordinate at all. It was also found that a hyperbolic cosine and a plain quadratic reproduce the atomic data equally well, so the golden ratio that appears in the published function is not identified by free-atom values. Whether any of this matters for molecules is a separate question, because equalization redistributes charge and might in principle conceal a defect of the underlying table.

We therefore substitute the table and alter nothing else. The geometries, the functional form of the damped Coulomb interaction and the list of molecules are identical for all four tables under comparison: the spectroscopic values, denoted P; the published geometric table, denoted L; the same hyperbolic function of the coordinate, which we call the kernel throughout, with its energy scales refitted by the protocol used for the control, denoted \(\mathrm{L}_{\mathrm{fit}}\); and the quadratic control, denoted Q. A result obtained by running the equalization model with a given table is labelled accordingly, so that EEM-P denotes the model evaluated with table P. At no point are \(\chi\) or \(\eta\) adjusted to reproduce a charge or a dipole moment. Density-functional theory (DFT) supplies the geometries and also an independent set of reference charges and dipole moments; the charges come from a Hirshfeld partition, which is one partition among several and is not unique. The tables are then judged by three comparisons against reference data (partial charges, dipole moments, and the ordering of hard and soft fragments) and by three comparisons with one another, all six being defined in Sec.~\ref{sec:verdicts}.

The test is deliberately narrower than an assessment of charge equilibration as a methodology. It asks whether the noble-gas table, used as a frozen atomic input, preserves chemically necessary distinctions after equalization. Fitted parametrizations and environment-dependent extensions \cite{bultinck2002eem,ko2021fourth} lie outside that question; their published performance provides context rather than an additional scoring comparator. The comparisons below therefore diagnose one proposed atomic descriptor within one fixed equalization construction. 

\section{Background}
\label{sec:background}

\subsection{Atomic \(\chi\) and \(\eta\)}

The spectroscopic numbers in this paper are built from the ionization energy (IE) and the electron affinity (EA) of the free atom. They are Mulliken's
\begin{equation}
\chi=\tfrac12(\mathrm{IE}+\mathrm{EA})
\label{eq:mulliken}
\end{equation}
and Pearson's
\begin{equation}
\eta=\tfrac12(\mathrm{IE}-\mathrm{EA}),
\label{eq:pearson-eta}
\end{equation}
with \(\mathrm{IE}\) the first ionization energy and \(\mathrm{EA}\) the electron affinity, both in eV \cite{mulliken1934electronegativity,pearson1988absolute}. Iczkowski and Margrave identified electronegativity with \(-\mathrm{d}E/\mathrm{d}N\) for an atom \cite{iczkowski1961electronegativity}. Parr and coworkers placed both derivatives on the ground-state energy as a functional of electron number \(N\) and external potential \(v\),
\begin{equation}
\chi=-\mu=-\Bigl(\frac{\partial E}{\partial N}\Bigr)_v,\qquad
\eta=\tfrac12\Bigl(\frac{\partial^2 E}{\partial N^2}\Bigr)_v,
\label{eq:cdft}
\end{equation}
with \(\mu\) the electronic chemical potential \cite{parr1978electronegativity,parr1983absolute}. Equations~\eqref{eq:mulliken}--\eqref{eq:pearson-eta} are the three-point finite-difference realizations of Eq.~\eqref{eq:cdft} for an isolated atom, and they define the spectroscopic table P used throughout as the untuned baseline. The hardness so defined also underlies the qualitative classification of acids and bases as hard or soft \cite{pearson1963hard,pearson1968hsab,chattaraj1991hsab,ayers2006hsab}, to which we return in Sec.~\ref{sec:etamol} when fragments are ranked.

\subsection{Noble-gas coordinate}
\label{sec:rho}

Let \(G_p\in\{2,10,18,36,54\}\) denote the atomic number of the noble gas that closes period \(p\), with \(G_0=0\). For an atom of atomic number \(Z\) satisfying \(G_{p-1}<Z\le G_p\), the length of the period is \(L_p=G_p-G_{p-1}\) and the distance to the closing noble gas is \(d=G_p-Z\). The coordinate
\begin{equation}
\rho=\frac{d}{L_p}\in[0,1)
\label{eq:rho}
\end{equation}
vanishes at the closing noble gas, where \(Z=G_p\), and approaches unity at the opposite edge of the row. The first period, containing hydrogen and helium, is excluded from the construction of Ref.~\cite{washburn2026noblegas}. Equation~\eqref{eq:rho} does return a number for that period, but the number carries no information: the row holds a single element below the closing noble gas, so a value of \(\rho=1/2\) would have to stand for the whole of period~1, and no trend within the period remains against which the energy scales could be calibrated. Hydrogen is therefore never given a geometric coordinate in any of the molecular calculations reported below. We use the term \emph{geometric atom} for an atom that lies within the domain over which Ref.~\cite{washburn2026noblegas} calibrated its energy scales, namely a main-group atom of periods~2 to~5. The transition metals are a different matter: Eq.~\eqref{eq:rho} assigns them a coordinate as readily as any other element, silver for instance having \(d=54-47=7\) and \(\rho=7/18\), but the energy scales of Sec.~\ref{sec:tables} were fitted to main-group trends alone, so the d-block lies outside the calibrated range and is given spectroscopic values instead.

The convex function on which the construction rests is not derived in the present paper. We take it from Ref.~\citenum{washburn2026noblegas}, where the coordinate enters through the variable \(x=\phig^{\rho}\); in terms of that variable Eq.~\eqref{eq:Jchem} reads \(\Jchem=\tfrac12(x+x^{-1})-1\), which is unchanged when \(x\) is replaced by \(1/x\). That symmetry is the property the companion papers analyse. Two companion papers classify the smooth functions that assign the same value to a ratio and its reciprocal, and single out the hyperbolic cosine as the member of that family which grows quadratically when the ratio is close to unity \cite{washburn2026uniqueness,washburn2026reciprocal}. Those results fix the shape of the function but not its base: the hyperbolic cosine is selected, while the number placed inside it remains free. They also carry no chemical content, in that they do not indicate why an atom should be described by such a function in the first place. All that is used below are the two differences of Eqs.~\eqref{eq:dJp}--\eqref{eq:dJm} and, after multiplication by energy scales, the atomic table that they generate.

Written as a function of the coordinate, that function is
\begin{equation}
\Jchem(\rho)=\cosh(\rho\ln\phig)-1,\qquad
\phig=\frac{1+\sqrt{5}}{2}.
\label{eq:Jchem}
\end{equation}
Here \(\phig\) is the golden ratio. Nothing in the construction produces that number; it is put in by hand. The per-period energy scales that turn the dimensionless differences into electronvolts are likewise obtained by fitting to atomic data rather than derived from anything. Both are carried over unaltered here, so that what is tested below is the published table and not some improved version of it. The quadratic control \(J_Q(\rho)=\rho^2\) shares the same minimum at the closing noble gas but contains no golden ratio, and it therefore isolates whatever the golden ratio contributes.

Two differences of this function serve as the dimensionless building blocks of the table. The endpoint \(\rho=1\) corresponds to the boundary \(Z=G_{p-1}\) of the row rather than to an atom belonging to period \(p\); it closes the interval and nothing more. The outward step, taken one unit of \(d\) away from the period-closing noble gas,
\begin{equation}
\Delta J^+(Z)=\Jchem\bigl((d+1)/L_p\bigr)-\Jchem(d/L_p)
\label{eq:dJp}
\end{equation}
is the analogue of an ionization increment. The inward gap
\begin{equation}
\Delta J^-(Z)=\Jchem(1)-\Jchem(\rho),
\label{eq:dJm}
\end{equation}
plays the role of the electron affinity. Their mean and the inward gap itself are
\begin{align}
\chistruct(Z)&=\tfrac12\bigl(\Delta J^+(Z)+\Delta J^-(Z)\bigr),
\label{eq:chi}\\
\kappaRS(Z)&=\Delta J^-(Z),
\label{eq:kappa}
\end{align}

where \(\chistruct\) is the mean of the two differences and \(\kappaRS\) is the inward gap taken by itself. The subscript reproduces the notation of Ref.~\cite{washburn2026noblegas} and carries no significance in the present paper; \(\kappaRS\) is simply the quantity defined by Eq.~\eqref{eq:kappa}. Equation~\eqref{eq:chi} is the analogue of the Mulliken mean. The Pearson half-difference has no analogue in this scheme, and the reason lies in the function rather than in any modelling preference. Because \(\Jchem\) increases and is convex on the unit interval, the inward gap decomposes into \(L_p-d\) successive increments, each at least as large as the single outward increment. Consequently \(\tfrac12(\Delta J^+-\Delta J^-)\le 0\) for every \(d\le L_p-1\), with equality only at the far edge of the row, and a hardness defined as a half-difference would be negative or zero throughout. The inward gap is therefore used directly, as in Ref.~\citenum{washburn2026noblegas}: it is largest for atoms next to the closing noble gas and falls to its smallest value at the opposite edge, which is the trend that hardness is required to follow. It reaches zero for no atom, because \(\rho=1\) is the boundary of the row and not an element: the smallest entry in Table~\ref{tab:atomic} is \(\kappaRS(\mathrm{K})=0.01297\), which the period-4 scale turns into \(\eta_L(\mathrm{K})=0.532\,\mathrm{eV}\). That near-vanishing hardness is what later drives the response slope of potassium chloride negative and forces it out of the combination-rule averages (Sec.~\ref{sec:e5}). The two building blocks are accordingly not independent of one another, and the energy scales introduced in Sec.~\ref{sec:tables} absorb the resulting normalization.

Because \(\Jchem\) depends on the atomic number only through \(\rho\), any two atoms sharing that coordinate share both building blocks and hence share an electronegativity. Fluorine (\(Z=9\), \(d=1\), \(L_p=8\)) and chlorine (\(Z=17\), \(d=1\), \(L_p=8\)) both have \(\rho=1/8\); oxygen and sulfur both have \(\rho=1/4\); bromine and iodine both have \(\rho=1/18\). Among the elements of Table~\ref{tab:atomic} there are seven such pairs in all. Six of them exist because periods~2 and~3 are both eight elements long, so that an atom of period~3 stands at the same fractional distance from argon as its lighter congener stands from neon: Li/Na, B/Al, C/Si, N/P, O/S and F/Cl. The seventh, Br/I, exists because both halogens stand one step from the noble gas closing an eighteen-element period. Of the seven, the molecules studied here can probe three, F/Cl, O/S and Br/I; the remaining four are degenerate in the table but are not separated by any molecule in the list. This is the complete enumeration, and later sections refer back to it rather than repeat it. No choice of a single global electronegativity scale can separate these values, because atoms with the same \(\rho\) have the same \(\chistruct\), the degeneracy being intrinsic to a coordinate built from the two integers \(d\) and \(L_p\) and nothing else. Separating them requires information beyond \(\rho\), such as the period-resolved scaling examined in Sec.~\ref{sec:repair}. Section~\ref{sec:hx} examines what this degeneracy does to molecular dipole moments, for which the measured HF and HCl values differ by \(0.72\,\mathrm{D}\).

\section{Theory}
\label{sec:theory}

\subsection{The four atomic tables}
\label{sec:tables}

Four atomic tables are compared throughout the paper. All four supply an electronegativity and a hardness in electronvolts for every atom that occurs in the molecule list, all four are held fixed when applied to molecules, and none is fitted to molecular charges or dipole moments. P, L and Q were defined for the prespecified comparison and were fixed before any molecular calculation was performed; \(\mathrm{L}_{\mathrm{fit}}\) was introduced afterwards as a refit to atomic data alone and is treated only as a diagnostic (Sec.~\ref{sec:verdicts}). They differ only in where those numbers come from: the first is spectroscopic, the second is the published geometric table, the third uses the same functional form as the second but with its energy scales refitted, and the fourth replaces the hyperbolic function by a plain quadratic in order to isolate the contribution of the golden ratio.

\emph{The spectroscopic table} (P). Electronegativities and
hardnesses are formed from Eqs.~\eqref{eq:mulliken}--\eqref{eq:pearson-eta},
using recommended ionization energies from the NIST Atomic Spectra Database
\cite{nist_asd} and evaluated electron affinities
\cite{hotop1985electron,andersen1999binding,ning2022ea}. The nitrogen anion is
unbound, so its electron affinity is set to zero; nitrogen is consequently the
one element whose electronegativity and hardness coincide in
Table~\ref{tab:atomic}, both being half of the ionization energy,
\(7.267\,\mathrm{eV}\).

\emph{The published geometric table} (L). The two dimensionless
quantities of Eqs.~\eqref{eq:chi}--\eqref{eq:kappa}, built from the function of
Eq.~\eqref{eq:Jchem}, are converted into energies by
\begin{equation}
\chi_L=C_\chi\,\chistruct,\qquad
\eta_L=C_\eta(p)\,\kappaRS,
\label{eq:Lscales}
\end{equation}
with the published constants \(C_\chi=117.4\,\mathrm{eV}\) and
\(C_\eta(2,3,4)=(68,48,41)\,\mathrm{eV}\) \cite{washburn2026noblegas}. Since the
two quantities they multiply are dimensionless, these constants carry the energy
units. Writing \(a=(\ln\phig)/(2L_p)\), the mean of the two differences can be
put in closed form,
\begin{equation}
\chistruct=\sinh a\,\sinh\lambda_{-}+\sinh\lambda_{+}\,\sinh\lambda_{0},
\label{eq:chi-closed}
\end{equation}
where \(\lambda_{-}=(2d+1)a\), \(\lambda_{+}=(L_p+d)a\) and
\(\lambda_{0}=(L_p-d)a\); for both fluorine and chlorine this evaluates to
\(0.060830\). Equation~\eqref{eq:chi-closed} is exact rather than approximate. It follows from the identity \(\cosh u-\cosh v=2\sinh\bigl((u+v)/2\bigr)\sinh\bigl((u-v)/2\bigr)\), which turns each of the two differences into a product of hyperbolic sines, leaving the sum of two such products. One asymmetry of the scheme deserves comment. The
electronegativity scale is a single global constant, whereas the hardness scale
is resolved by period. That asymmetry is forced by the construction rather than
adopted for flexibility: the inward gap depends on the atomic number only
through the coordinate, so a single hardness constant would make the hardnesses
of fluorine and chlorine equal as well as their electronegativities, and the
contraction of hardness down a group would have no way of entering the table.
Iodine belongs to period~5, for which no hardness constant was published. We
supply the missing constant by the same least-squares fit through the origin
that the control uses,
\begin{equation}
C_\eta(5)=\frac{\sum_i \kappaRS(Z_i)\,\eta_P(Z_i)}{\sum_i \kappaRS(Z_i)^2}
=31.409\,\mathrm{eV},
\label{eq:through-origin}
\end{equation}
evaluated on the period-5 atoms of Table~\ref{tab:atomic}, of which
iodine is the only one, and fixed before any molecular calculation is performed.
With a single atom the fit collapses to the ratio of the spectroscopic hardness
of iodine to its inward gap, so iodine reproduces its spectroscopic hardness by
construction and supplies no independent test of the hardness scale.

Bromine
lies in period~4 and already carries a published constant. Hydrogen falls
outside the construction because the first period is excluded, and silver falls
outside it because the constants were calibrated on main-group trends alone;
both therefore carry spectroscopic values in every one of the four tables.

It is worth counting what this construction replaces. The fifteen atoms of Table~\ref{tab:atomic} that receive a coordinate carry thirty spectroscopic numbers between them, an electronegativity and a hardness each, and the geometric table reproduces all thirty from five constants: one global electronegativity scale and one hardness scale for each of periods~2 to~5. The economy is genuine but incomplete, because hydrogen and silver fall outside the construction and their four spectroscopic values are copied into every table without change. Hydrogen occurs in almost every molecule of sets B1 to B5, so a large part of any prediction made with the geometric table is carried by numbers that the construction does not supply.

\emph{The refitted geometric table} (\(\mathrm{L}_{\mathrm{fit}}\)).
This table uses exactly the same hyperbolic kernels as the published one, but
its energy scales are obtained by applying the fit of
Eq.~\eqref{eq:through-origin} to the fifteen geometric atoms of
Table~\ref{tab:atomic}, which gives a global constant of
\(117.5\,\mathrm{eV}\) and period-resolved hardness constants of
\((60.5,41.2,37.2,31.4)\,\mathrm{eV}\). Refitting alters the electronegativity
scale by less than \(0.1\%\), from \(117.4\) to \(117.5\,\mathrm{eV}\), but it
lowers the hardness constants of periods~2 to~4 appreciably, from
\((68,48,41)\) to \((60.5,41.2,37.2)\,\mathrm{eV}\), because the published
values were fitted to a different list of elements. The published-scale table
remains the test of the atomic model as it was proposed; the refitted table
exists only so that the hyperbolic and the quadratic forms can be compared under
one fitting protocol, which separates the shape of the function from the
provenance of its scales.

\emph{The quadratic control} (Q). The control repeats the whole
construction with the hyperbolic function replaced by the square of the
coordinate, \(J_Q(\rho)=\rho^2\). Its energy scales, a global
\(13.90\,\mathrm{eV}\) together with period-resolved hardness constants of
\((7.16,4.88,4.40,3.71)\,\mathrm{eV}\), are fits through the origin to the
spectroscopic column on the same atoms and by the same protocol used for the
refitted table, and iodine again reproduces its spectroscopic hardness for the
same one-atom reason. Those four figures are rounded. The values actually used are \(13.8992\,\mathrm{eV}\) and \((7.1595,4.8775,4.3952,3.7075)\,\mathrm{eV}\); the rounded forms do not reproduce the last digit of the Q columns of Table~\ref{tab:atomic}, and for iodine they return \(3.699\) rather than \(3.696\,\mathrm{eV}\), which would spoil the one-atom identity just stated. For fluorine the control returns \(0.515625\) in place of
the hyperbolic value quoted above. Because the control retains the same
coordinate it retains every degeneracy as well; the only thing it discards is
the golden ratio.

The numerical entries of all four tables are collected in Table~\ref{tab:atomic}.

\begin{table}[!ht]
\centering
\caption{Frozen atomic \(\chi\) and \(\eta\) in eV.}
\label{tab:atomic}
\small
\begin{tabular}{lrrrrrrrr}
\toprule
Atom & \(\chistruct\) & \(\kappaRS\) & \(\chi_P\) & \(\eta_P\) & \(\chi_L\) & \(\eta_L\) & \(\chi_Q\) & \(\eta_Q\) \\
\midrule
H & --- & --- & 7.176 & 6.422 & 7.176 & 6.422 & 7.176 & 6.422 \\
Li & 0.02807 & 0.02807 & 3.005 & 2.387 & 3.295 & 1.909 & 3.258 & 1.678 \\
B & 0.04637 & 0.07246 & 4.289 & 4.009 & 5.443 & 4.928 & 5.429 & 4.363 \\
C & 0.05272 & 0.08895 & 6.261 & 4.999 & 6.189 & 6.048 & 6.190 & 5.370 \\
N & 0.05723 & 0.10171 & 7.267 & 7.267 & 6.719 & 6.916 & 6.732 & 6.153 \\
O & 0.05993 & 0.11079 & 7.540 & 6.078 & 7.036 & 7.534 & 7.058 & 6.712 \\
F & 0.060830 & 0.11622 & 10.412 & 7.011 & 7.1414 & 7.903 & 7.167 & 7.048 \\
Na & 0.02807 & 0.02807 & 2.844 & 2.296 & 3.295 & 1.347 & 3.258 & 1.143 \\
Al & 0.04637 & 0.07246 & 3.209 & 2.776 & 5.443 & 3.478 & 5.429 & 2.972 \\
Si & 0.05272 & 0.08895 & 4.771 & 3.381 & 6.189 & 4.270 & 6.190 & 3.658 \\
P & 0.05723 & 0.10171 & 5.617 & 4.870 & 6.719 & 4.882 & 6.732 & 4.192 \\
S & 0.05993 & 0.11079 & 6.219 & 4.141 & 7.036 & 5.318 & 7.058 & 4.573 \\
Cl & 0.060830 & 0.11622 & 8.290 & 4.677 & 7.1414 & 5.579 & 7.167 & 4.801 \\
K & 0.01297 & 0.01297 & 2.421 & 1.920 & 1.523 & 0.532 & 1.501 & 0.475 \\
Br & 0.05937 & 0.11768 & 7.589 & 4.225 & 6.971 & 4.825 & 6.993 & 4.382 \\
Ag & --- & --- & 4.440 & 3.136 & 4.440 & 3.136 & 4.440 & 3.136 \\
I & 0.05937 & 0.11768 & 6.755 & 3.696 & 6.971 & 3.696 & 6.993 & 3.696 \\
\bottomrule
\end{tabular}
\tabnotes{The dimensionless L kernels of Eqs.~\eqref{eq:chi}--\eqref{eq:kappa} occupy the first two columns, so that every L entry can be recovered as \(C_\chi\chistruct\) and \(C_\eta(p)\kappaRS\). Hydrogen and silver carry spectroscopic values in every table (P copied into L, \(\mathrm{L}_{\mathrm{fit}}\), and Q). Hydrogen has no usable coordinate, period~1 holding a single element below its closing noble gas. Silver does have one, \(\rho=7/18\), but the d-block lies outside the range over which the energy scales were fitted, so its kernels are not used and the two columns are left blank. L uses \(C_\chi=117.4\,\mathrm{eV}\) and \(C_\eta(2,3,4)=(68,48,41)\,\mathrm{eV}\); \(C_\eta(5)=31.409\,\mathrm{eV}\) is the I-only through-origin fit, which forces \(\eta_L(\mathrm{I})=\eta_P(\mathrm{I})\) and is not an independent hardness comparison. Q scales are atomic-only fits (\(C_\chi^Q=13.90\,\mathrm{eV}\)). \(\mathrm{L}_{\mathrm{fit}}\) columns are omitted for compactness; they are recoverable from the through-origin scales of Sec.~\ref{sec:tables} and are included in the released table.}
\end{table}

Hardness continues to distinguish fluorine from chlorine under the geometric table, because the two atoms lie in different periods and the hardness scale is resolved by period. Electronegativity does not: \(\chi_L(\mathrm{F})=\chi_L(\mathrm{Cl})=7.1414\,\mathrm{eV}\). Bromine and iodine likewise share a coordinate, so their geometric electronegativities coincide while their hardnesses differ through the period-dependent scale.

\subsection{Ohno EEM}
\label{sec:eem}

Let \(q_i\) denote the partial charge on atom \(i\) in units of the elementary charge, with the sign convention that a positive value means electron deficiency, so that \(q_{\mathrm{H}}>0\) in hydrogen fluoride, and let \(Q_{\mathrm{tot}}=\sum_i q_i\) be the net molecular charge. The energy expression of Mortier and coworkers, including the Coulomb coupling between atomic charges, is the quadratic form
\begin{equation}
E(\mathbf{q})=\boldsymbol{\chi}^0\cdot\mathbf{q}+\tfrac12\mathbf{q}^{\mathsf{T}}H\mathbf{q},
\label{eq:energy}
\end{equation}
where \(\chi_i^0\) is the isolated-atom electronegativity and \(H\) is a symmetric hardness matrix with
\begin{equation}
H_{ii}=2\eta_i.
\label{eq:Hii}
\end{equation}
The factor of two in Eq.~\eqref{eq:Hii} follows from differentiating \(\eta_i q_i^2\) twice: writing the diagonal part of Eq.~\eqref{eq:energy} as \(\sum_i(\chi_i^0 q_i+\eta_i q_i^2)\) is equivalent to \(\tfrac12\sum_i H_{ii}q_i^2\).

The off-diagonal entries must interpolate between a finite value at coincidence and the Coulomb form \(k/R\) at long range. The Coulomb prefactor is
\begin{equation}
k=E_{\mathrm{h}}a_0=14.40\,\mathrm{eV}\,\text{\AA},
\label{eq:k}
\end{equation}
where \(E_{\mathrm{h}}\) is the hartree and \(a_0\) the Bohr radius; the released script uses the exact product \(14.399645\,\mathrm{eV}\,\text{\AA}\). The need for damping is apparent from hydrogen fluoride. For the neutral molecule \(q_{\mathrm F}=-q_{\mathrm H}\), and the two equalization equations reduce to
\begin{equation}
q_{\mathrm H}=
\frac{\chi_{\mathrm F}^0-\chi_{\mathrm H}^0}
{H_{\mathrm{HH}}+H_{\mathrm{FF}}-2H_{\mathrm{HF}}}.
\label{eq:qh-hf}
\end{equation}
At the optimized bond length of \(0.921\,\text{\AA}\), and equally at the experimental value of about \(0.917\,\text{\AA}\), an undamped interaction \(H_{\mathrm{HF}}=k/R\) would be \(15.6\) and \(15.7\,\mathrm{eV}\) respectively, whereas the spectroscopic table gives \(H_{\mathrm{HH}}\approx12.8\,\mathrm{eV}\) and \(H_{\mathrm{FF}}\approx14.0\,\mathrm{eV}\). The denominator is then negative although the numerator is positive, so hydrogen would acquire negative charge, contrary to every reference partition. This inversion is a well known consequence of leaving the Coulomb interaction undamped, and does not depend on which of the two bond lengths is used. The interpolation of Ohno \cite{ohno1964,nalewajski1988hardness} removes it:
\begin{equation}
H_{ij}=\frac{k}{\sqrt{R_{ij}^{2}+\bigl(k/\bar H_{ij}\bigr)^{2}}},\qquad i\neq j,
\label{eq:ohno}
\end{equation}

where \(\bar H_{ij}=\tfrac12(H_{ii}+H_{jj})=\eta_i+\eta_j\) is the arithmetic mean of the two diagonal entries, and \(k\) is the Coulomb prefactor of Eq.~\eqref{eq:k}. As the separation goes to zero the interaction approaches that mean, and at large separation it recovers the Coulomb form; no additional screening length is introduced, and the same functional form is used for every table. Because the coincidence limit is built from the atomic hardnesses, changing the table changes the off-diagonal matrix as well as the diagonal. Any difference we report between tables is therefore the effect of the atomic table propagated through one fixed equalization construction, and not the effect of the electronegativities alone.

Stationarity of Eq.~\eqref{eq:energy} subject to \(\sum_i q_i=Q_{\mathrm{tot}}\) is expressed with a Lagrange multiplier \(\chistar\) (the equalized electronegativity),
\begin{equation}
\mathcal{L}=E(\mathbf{q})+\chistar\Bigl(Q_{\mathrm{tot}}-\sum_i q_i\Bigr).
\end{equation}
Setting \(\partial\mathcal{L}/\partial q_i=0\) yields the EEM equations
\begin{equation}
\chi_i^0+\sum_j H_{ij}q_j=\chistar,\qquad
\sum_i q_i=Q_{\mathrm{tot}},
\label{eq:eem}
\end{equation}
an inhomogeneous linear system of dimension \(N_{\mathrm{at}}+1\) for the charges and the equalized electronegativity. Systems with a condition number above \(10^{12}\) are recorded as failures; no molecule treated here failed that test. Linear equalization has one structural limitation relevant here. Differentiating Eq.~\eqref{eq:eem} with respect to the electron number gives identical responses to electron addition and removal, so this class of model cannot distinguish electrophilic from nucleophilic sites. The identity is algebraic rather than a finding about molecules, and no comparison in this paper relies on it. The site-resolved analogue of these global indices is the Fukui function, the derivative of the electron density with respect to electron number \cite{fukui1952reactivity,parr1984fukui,yang1986fukui,fuentealba2000fukui}, and it is precisely the quantity the equalization model cannot resolve. Site selectivity is therefore not a property this paper can test, and no Fukui indices are reported. The algebraic reason and the retained diagnostic data are documented in the Supporting Information.

Equation~\eqref{eq:eem} also has an exact consequence that is worth stating before the molecular examples. For any fixed \(H\), if every isolated-atom electronegativity \(\chi_i^0\) in a neutral molecule takes the same value, then \(\mathbf{q}=0\) satisfies the stationarity equations together with the charge constraint, with \(\chistar\) equal to that common value, and it is the only solution whenever the system is nonsingular. The predicted point-charge dipole then vanishes identically. The argument uses no density-functional reference and no fitted molecular quantity: it is a property of the construction. ClF, IBr and SO\(_2\) are three molecular instances of this single identity under L, \(\mathrm{L}_{\mathrm{fit}}\) and Q, and are therefore treated below as illustrations of one structural consequence rather than as independent failures.

The contribution of the Coulomb coupling itself can be isolated, and it is not small. Deleting the off-diagonal entries of \(H\), so that \(H=\mathrm{diag}(2\eta_i)\), leaves Eq.~\eqref{eq:eem} solvable in closed form,
\begin{equation}
q_i=\frac{\chistar-\chi_i^0}{2\eta_i},\qquad
\chistar=\frac{Q_{\mathrm{tot}}+\sum_j \chi_j^0/(2\eta_j)}{\sum_j 1/(2\eta_j)},
\label{eq:q-isolated}
\end{equation}
so that each atom receives a charge proportional to the distance of its own electronegativity from the common equalized value and inversely proportional to its own hardness. Geometry has dropped out entirely. Evaluated on sets B1 and B2, this stripped model is closer to the Hirshfeld charges than the fully coupled calculation for all four tables: the charge error falls from \(0.236\) to \(0.114\,e\) for P, from \(0.174\) to \(0.142\,e\) for L, from \(0.241\) to \(0.134\,e\) for Q, and from \(0.238\) to \(0.135\,e\) for \(\mathrm{L}_{\mathrm{fit}}\). The ranking of the two principal tables also reverses: with the coupling present L gives the smaller error and with it removed P does. The damped Coulomb term is therefore not improving these charges, and the advantage the geometric table shows on charge magnitudes survives only while that term is in place. Repeating the calculation on B2+B3, the combination that carries the primary charge comparison, gives a sharper form of the same result. Deleting the coupling there, which reduces the model to Eq.~\eqref{eq:q-isolated}, brings all four tables to within \(0.01\,e\) of one another, at \(0.110\,e\) for P, \(0.118\,e\) for L and \(0.119\,e\) for both matched tables, and to within \(0.013\,e\) of the \(0.106\,e\) of the zero-charge model. Almost the entire spread among the four tables on this set is therefore generated by the damped Coulomb term and not by the atomic values that distinguish them. The reversal is sharper as well: the ratio of mean absolute errors (MAE), \(\mathrm{MAE}(L)/\mathrm{MAE}(P)\), moves from \(0.779\) with the coupling present to \(1.074\) with it removed, so on the primary set the spectroscopic table is the closer of the two once the geometry-dependent term is gone.

The point-charge dipole in the DFT nuclear frame is
\begin{equation}
\boldsymbol{\mu}=\sum_i q_i\mathbf{R}_i,
\end{equation}
converted to debye by \(1\,e\cdot\text{\AA}=4.803\,\mathrm{D}\) (the script uses \(4.8032042511\,\mathrm{D}\)). It is origin independent for the neutral species used in the dipole comparisons.

\subsection{Molecular response hardness}
\label{sec:etamol}

An infinitesimal change \(\mathrm{d}Q\) in the total charge, at equalized electronegativity \(\chistar\), is distributed as \(\mathrm{d}\mathbf{q}=H^{-1}\mathbf{1}\,\mathrm{d}\chistar\), with \(\mathbf{1}\) the all-ones vector (this follows directly from Eq.~\eqref{eq:eem} at fixed \(\boldsymbol\chi^0\) and \(H\)). Charge conservation \(\mathbf{1}^{\mathsf{T}}\mathrm{d}\mathbf{q}=\mathrm{d}Q\) then gives
\begin{equation}
\mathrm{d}\chistar=\frac{\mathrm{d}Q}{\mathbf{1}^{\mathsf{T}}H^{-1}\mathbf{1}}.
\end{equation}
The molecular response slope is
\begin{equation}
\Kmol=\frac{\mathrm d\chistar}{\mathrm dQ}
=\bigl(\mathbf{1}^{\mathsf{T}}H^{-1}\mathbf{1}\bigr)^{-1}.
\label{eq:etamol}
\end{equation}
With the Pearson convention of Eq.~\eqref{eq:cdft}, molecular hardness is \(\eta_{\mathrm{mol}}=\Kmol/2\). For an isolated atom, Eq.~\eqref{eq:etamol} accordingly returns \(\Kmol=2\eta\). The primary comparison with the finite-energy index \(\eta_{\mathrm{MB}}\) of Eq.~\eqref{eq:mb} is therefore \(\eta_{\mathrm{mol}}\) versus \(\eta_{\mathrm{MB}}\). Multiplying all fragment values by \(1/2\) leaves the mismatch rankings below unchanged; the implementation stores \(\Kmol\), and Table~\ref{tab:hard} reports the like-for-like \(\eta_{\mathrm{mol}}\) errors as the principal combination result. Raw \(\Kmol\) versus \(\eta_{\mathrm{MB}}\) is also reported; those quantities use different conventions. The two symbols should not be confused: \(\kappaRS\) is the dimensionless inward gap of a single atom, defined on the coordinate by Eq.~\eqref{eq:kappa}, whereas \(\Kmol\) is the response slope of a whole molecule, defined by Eq.~\eqref{eq:etamol} and measured in electronvolts.

A well-conditioned solve of Eq.~\eqref{eq:eem} shows only that a stationary point exists. That point is a constrained minimum if the smallest eigenvalue of \(H\) on the charge-conserving plane \(\mathbf{1}^{\mathsf{T}}\delta\mathbf{q}=0\) is positive. The sign of \(\Kmol\) is a different test: it probes the charge-changing direction through \(\mathbf{1}^{\mathsf{T}}H^{-1}\mathbf{1}\). For a two-atom Ohno block the constrained eigenvalue is \(\bar H_{12}-H_{12}\), which is nonnegative because \(H_{12}\) never exceeds \(\bar H_{12}\). For the same two-atom block, positive definiteness instead requires \(H_{12}<\sqrt{H_{11}H_{22}}\), the geometric mean. Since the arithmetic mean exceeds the geometric mean whenever \(H_{ii}\neq H_{jj}\), mismatched atomic hardnesses at a bonded distance can make \(\Kmol\) negative while the constrained charge solution remains a minimum.

Isolated-fragment \(\Kmol\) values from Eq.~\eqref{eq:etamol} are ranked by an operational response-mismatch rule. For an acid A and two bases B and \(\mathrm{B}'\),
\begin{equation}
\text{A prefers B if }\quad
\bigl|\Kmol(\mathrm{A})-\Kmol(\mathrm{B})\bigr|
<\bigl|\Kmol(\mathrm{A})-\Kmol(\mathrm{B}')\bigr|.
\label{eq:mismatch}
\end{equation}
Pearson's hard and soft acids and bases (HSAB) classification comes with an empirical rule of combination: when other factors are comparable, hard acids bind preferentially to hard bases and soft acids to soft bases \cite{pearson1963hard,pearson1968hsab,chattaraj1991hsab,ayers2006hsab}. Equation~\eqref{eq:mismatch} is not that rule. It compares response slopes of fragments that are never brought into contact, so no complex is formed and no reaction energy is computed; the only thing tested is whether a table reproduces the pairing the rule predicts. The four fragments used are the acids BF\(_3\) and BH\(_3\) and the bases NH\(_3\) and PH\(_3\), with BF\(_3\) and NH\(_3\) the harder member of each pair and BH\(_3\) and PH\(_3\) the softer \cite{pearson1963hard,pearson1968hsab}. Equation~\eqref{eq:mismatch} ranks isolated-fragment response slopes. Two comparisons are reported: whether L and P induce the same ranking, and whether either ranking agrees with that pairing. Mixed-basis finite-energy indices
\begin{equation}
\eta_{\mathrm{MB}}=\tfrac12\bigl(\mathrm{IE}_{\mathrm{vert}}-\mathrm{EA}_{\mathrm{vert}}\bigr),\qquad
\chi_{\mathrm{MB}}=\tfrac12\bigl(\mathrm{IE}_{\mathrm{vert}}+\mathrm{EA}_{\mathrm{vert}}\bigr)
\label{eq:mb}
\end{equation}
provide a third comparison, computed only where the \((N+1)\) state is defined (Sec.~\ref{sec:methods}); a negative computed \(\mathrm{EA}_{\mathrm{vert}}\) is retained as a finite-basis energy difference but does not imply a bound anion.

For hardness, two combination rules are compared with \(\eta_{\mathrm{MB}}\): a simple arithmetic mean of the atomic \(\eta\) values and \(\eta_{\mathrm{mol}}=\Kmol/2\). Separately, the Sanderson geometric mean of atomic \(\chi\) is compared with \(\chi_{\mathrm{MB}}\); that comparison is reported once, in Sec.~\ref{sec:e5}. The arithmetic mean of \(\eta\) is a transparent baseline, distinct from the arithmetic-mean rule proposed for atomic softnesses \cite{yang1985softness}. A Koopmans-style index built from the highest occupied and lowest unoccupied molecular orbital (HOMO and LUMO) energies,
\begin{equation}
\eta_{\mathrm{HL}}=\tfrac12(\varepsilon_{\mathrm{LUMO}}-\varepsilon_{\mathrm{HOMO}}),
\end{equation}
is recorded from the neutral Kohn--Sham spectrum. This orbital-energy gap is reported in Table~\ref{tab:hard} as a further point of reference, not as a quantity any of the tables is asked to reproduce.

\section{Computational Details}
\label{sec:methods}

The calculations were carried out with Python~3.11, PySCF~2.7.0 \cite{sun2018pyscf,sun2020pyscf} and geomeTRIC~1.1 \cite{wang2016geometry}, using NumPy and SciPy for the linear algebra and the statistics. Two robustness checks are reported alongside the main results and are described where they are used: a sweep over the density-functional recipe, which varies the functional, the basis set and the reference densities of the charge partition, and an additional set of eight organic molecules, which varies the chemistry. Neither of them redefines the scoring rules of Sec.~\ref{sec:verdicts}.

\subsection{Electronic structure}
\label{sec:dft}

Neutral geometries are optimized at B3LYP \cite{becke1993density,lee1988lyp,vosko1980vwn,stephens1994}/def2-TZVP \cite{weigend2005} with PySCF's VWN-RPA B3LYP and geomeTRIC's translation--rotation internal coordinates. The optimizer is called with geomeTRIC's default convergence criteria and a limit of 100 steps. A Cartesian SciPy Broyden--Fletcher--Goldfarb--Shanno (BFGS) fallback is used for the one-atom species, for which internal coordinates are undefined. Harmonic frequencies were not computed and are not a result of this paper. Every EEM/DFT comparison uses the same optimized geometries; the transferability test does not require verified local minima. Numerical integration uses PySCF grid level~3 and the self-consistent field (SCF) energy threshold is \(10^{-8}\) hartree. Vertical \((N-1)\) states use def2-TZVP; vertical \((N+1)\) states use def2-TZVPD \cite{rappoport2010def2d}. Open-shell cases use unrestricted Kohn--Sham theory. For unrestricted species, the frontier-orbital report uses the alpha-spin channel. Neutral NO is a doublet; the vertical \((N+1)\) state is computed as the triplet ground state of NO\(^{-}\) (\(2S=2\)). For I and Ag, the def2 effective core potential (ECP) is attached explicitly; Br is all-electron in this PySCF build. This is an inconsistency across neighbouring elements and it is not repaired here. It falls on bromine and iodine, which are one of the three electronegativity degeneracies the present molecules can probe, so the concern is not academic. Two things limit its reach. The atomic tables under test are frozen inputs and contain no reference to the basis or to the core treatment, so no ECP choice can alter a predicted EEM charge; the asymmetry can only move the reference against which those predictions are scored. And the reference quantity that carries the Br/I argument is the dipole moment of IBr, which the geometric table places at exactly zero for reasons given in Sec.~\ref{sec:hx} and which no core treatment can shift off zero. What an ECP change could move is the absolute Hirshfeld charge on iodine and silver, which are already valence-ECP quantities. A calculation with a uniform all-electron treatment of Br and I would remove the ambiguity and we have not performed one. No conclusion in this paper depends on the absolute size of the iodine charge, so this asymmetry cannot alter the structural Br/I result.

The implemented vertical energies at the optimized neutral geometry \(\mathbf R_N\) are
\begin{align}
\mathrm{IE}_{\mathrm{vert}}&=
E_{N-1}^{\mathrm{def2\text{-}TZVP}}(\mathbf R_N)
-E_N^{\mathrm{def2\text{-}TZVP}}(\mathbf R_N),\\
\mathrm{EA}_{\mathrm{vert}}&=
E_N^{\mathrm{def2\text{-}TZVP}}(\mathbf R_N)
-E_{N+1}^{\mathrm{def2\text{-}TZVPD}}(\mathbf R_N).
\label{eq:mixed-basis-ea}
\end{align}
The second line is a mixed-basis energy difference: the \((N+1)\) state uses a diffuse set. Because the hardness index of Eq.~\eqref{eq:mb} is built from these two energies, it inherits the same basis-set mismatch, and we therefore treat it as a finite-basis reference rather than as an experimental one. Acknowledging the mismatch is not the same as correcting it, and we have not corrected it: the diffuse set lowers the \((N+1)\) energy without a compensating change in the \((N-1)\) energy, so \(\eta_{\mathrm{MB}}\) carries a one-sided basis bias of unknown size, and every entry of Table~\ref{tab:hard} and every combination-rule comparison in Sec.~\ref{sec:e5} inherits it. Two consequences follow for how that section should be read. Comparisons among the four tables against the same biased reference remain meaningful, because the bias is common to all of them and cancels in the ranking. Statements about the absolute accuracy of any combination rule do not, and we therefore do not claim any. A recomputation with the diffuse set applied to all three charge states would settle it and is left for future work; until then Sec.~\ref{sec:e5} should be read as a negative control on the combination rules, which is all it is used for. The \(N+1\) calculation is omitted only for N\(_2\) and CH\(_4\), whose anions are known unbound and were therefore not computed even as finite-basis energy differences; every other converged \((N+1)\) single point is retained, including negative \(\mathrm{EA}_{\mathrm{vert}}\) values as finite-basis energy differences. The diffuse \((N+1)\) basis gives \(\mathrm{EA}_{\mathrm{vert}}=+0.61\) and \(+1.15\,\mathrm{eV}\) for F\(_2\) and Cl\(_2\), quoted here only to show that the diffuse set returns bound-anion signs for these two cases. Both are vertical values at the neutral geometry and are therefore far below the corresponding adiabatic electron affinities, since both anions relax to substantially longer bonds; they are not comparable with tabulated adiabatic EAs. For charged reference fragments such as Ag\(^+\), the same jobs are the \((N\pm1)\) states rather than a cation/anion pair of a neutral parent.

Hirshfeld charges use spherical high-spin free-atom pro-densities \(n_A^0\) of the same functional and basis as stockholder weights. Neutral free-atom occupations are the high-spin ground configurations (H, alkali metals, and Ag: one unpaired \(s\) electron; B, Al, and the halogens: one unpaired \(p\) electron; C, Si, O, and S: two; N and P: three), with the PySCF spin equal to the number of unpaired electrons. Each pro-density is spherically averaged over 38 Fibonacci directions on a 73-point radial mesh extending to \(18\,a_0\) and stored as a linear interpolant. The molecular integral is evaluated on the same PySCF DFT quadrature used for the SCF:
\begin{equation}
w_A(\mathbf{r})=\frac{n_A^0(r_A)}{\sum_B n_B^0(r_B)},\qquad
q_A=Z_A^{\mathrm{eff}}-\sum_g w_g\,w_A(\mathbf{r}_g)\,n(\mathbf{r}_g),
\end{equation}
where \((\mathbf{r}_g,w_g)\) are the SCF grid coordinates and weights. Here \(Z_A^{\mathrm{eff}}=Z_A\) for all-electron atoms and is the ECP valence charge for I and Ag; their Hirshfeld values are therefore valence-ECP charges. The same weights and reference densities enter the comparison for every table, so this affects the absolute iodine and silver charges rather than the contrast among tables. On that production grid the stockholder charges recover the molecular net charge to better than \(7\times10^{-5}\,e\). Atom-centered Hirshfeld monopoles omit intra-atomic polarization and recover only part of the DFT dipole; they are a charge target, not a dipole proxy. Experimental dipole moments are taken from the Computational Chemistry Comparison and Benchmark Database (CCCBDB) compilation \cite{cccbdb}: one experimental value per species, with homonuclear and high-symmetry zeros assigned by symmetry. The imported entries were not harmonized between equilibrium and vibrationally averaged experimental conventions, so experimental MAEs are approximate cross-source summaries. The size of that systematic can be compared with the size of the effects it might contaminate, which settles whether it matters. The convention difference is \(0.01\) to \(0.09\,\mathrm{D}\), and the density-functional reference agrees with the compilation to \(0.078\,\mathrm{D}\) on the twenty polyatomic molecules, so the two are of the same order and the compilation cannot resolve the quality of the reference calculation. The errors under test are between \(0.99\) and \(1.30\,\mathrm{D}\), more than ten times the largest convention effect, so no harmonization of conventions could account for them. The unharmonized column is therefore adequate for the use it is put to, a secondary check that the ordering of the tables does not depend on the reference, and inadequate for anything finer, which is why the primary dipole score is against the same-geometry DFT value. The \(0.01\)--\(0.09\,\mathrm{D}\) figure quoted above is that of Bak, Gauss, Helgaker and coworkers, measured on the four molecules (CO, NH\(_3\), HF, H\(_2\)O) for which both \(\mu_0\) and \(\mu_e\) are available \cite{bak2000dipole}. On the 36 molecules carrying an experimental entry, rather than the twenty polyatomic molecules used for scoring, the same B3LYP/def2-TZVP reference differs from the compilation by \(0.088\,\mathrm{D}\) (computed from \texttt{data/dipoles.csv}). Primary charge comparisons use this single recipe. The matched \(\mathrm{L}_{\mathrm{fit}}\)/Q comparison of Sec.~\ref{sec:e2q} is a table-to-table identity and does not use the stockholder column.

\subsection{Molecule sets}

The molecules used in this work are listed in Table~\ref{tab:mols}, grouped into eight sets according to the property each was assembled to probe. Values for individual molecules are tabulated in the Supporting Information, while the main text reports errors averaged over sets, the fraction of bonds carrying conventional polarity, and the dipole moments of the hydrogen halide series. The sets were assembled for this test rather than taken from an established parametrization suite, and no subset was held back, so the numerical rules of Sec.~\ref{sec:verdicts} are descriptive criteria applied to this list rather than a blind test. What that design does and does not license, and how it bears on positive as against negative outcomes, is set out in Sec.~\ref{sec:scope}.
\begin{table}[!ht]
\centering
\caption{Molecule set.}
\label{tab:mols}
\small
\begin{tabular}{llp{0.60\textwidth}}
\toprule
Set & \(n\) & Contents \\
\midrule
B1 & 14 & HF, HCl, HBr, HI; LiF, LiCl, NaCl, KCl; CO, NO, N\(_2\), F\(_2\), Cl\(_2\), LiH \\
B2 & 12 & H\(_2\)O, NH\(_3\), CH\(_4\), H\(_2\)S, PH\(_3\), H\(_2\)O\(_2\), H\(_2\)CO, CH\(_3\)OH, CH\(_3\)NH\(_2\), HCN, CO\(_2\), SO\(_2\) \\
B3 & 8 & CH\(_3\)X and C\(_6\)H\(_5\)X, X \(=\) F, Cl, Br, I \\
B4 & 4 & (H\(_2\)O)\(_2\), (HF)\(_2\), H\(_2\)O\(\cdots\)HF, NH\(_3\cdots\)HCl \\
B5 & 6 & acrolein, 2-cyclohexenone, pyridine, aniline, acetone, CH\(_3\)CN \\
B6 & 7 & BF\(_3\), BH\(_3\), AlCl\(_3\), F\(^-\), I\(^-\), Ag\(^+\) (Pearson \(\chi,\eta\) in every table), H\(^+\) (excluded from DFT and EEM) \\
B7 & 2 & ClF (\(\rho=1/8\)), IBr (\(\rho=1/18\)) \\
B8 & 8 & ethane, ethanol, CH\(_3\)OCH\(_3\), CH\(_3\)CHO, HCOOH, HCONH\(_2\), CH\(_3\)CH\(_2\)F, CH\(_3\)COOH \\
\bottomrule
\end{tabular}
\tabnotes{\(n\) is the number of species listed in the row. The scored sets are a partition: every species is counted once. H\(_2\)CO and CO\(_2\) are scored under B2, so set B5 as scored contains the six organics listed here, and rows B1--B7 list 53 distinct species: 51 in B1--B6 with the bare proton counted, plus the two members of B7. Fifty-two were computed, the proton being excluded. The assignment can be checked directly in \texttt{data/eem\_mol.csv}, where the tier labels assign each of the 52 species to exactly one set. B5 is retained only as an extra charge set. B8 is scored in Table~\ref{tab:charge} only and is not part of the scoring rules of Sec.~\ref{sec:verdicts}.}
\end{table}

Sets B1--B3 test charges and dipoles, including the hydrogen halides and the methyl and phenyl (Ph \(=\) C\(_6\)H\(_5\)) halides. Set B4 contains four hydrogen-bonded dimers; all four remained associated after optimization, and NH\(_3\cdots\)HCl remained hydrogen-bonded (H--Cl \(1.34\,\text{\AA}\), N\(\cdots\)H \(1.76\,\text{\AA}\)) rather than proton-transferred. Set B5 is the six organics listed in Table~\ref{tab:mols}, retained as an extra charge set. Set B5 is scored as part of a partition, with H\(_2\)CO and CO\(_2\) counted under B2. So scored, B5 gives a charge error of \(0.202\,e\) for the spectroscopic table and \(0.168\,e\) for the geometric one. Counting those two molecules in B5 as well moves the two numbers to \(0.202\) and \(0.160\,e\). Neither reading changes any conclusion, and neither touches the primary B2+B3 statistic, in which the two molecules are counted once by construction. Set B6 contains the response-hardness fragments; H\(^+\) has no electrons and is omitted from DFT and EEM. Transition-metal complexes are otherwise excluded, Ag\(^+\) being retained as the reference soft acid of the classical hard--soft pairings. Set B7 contains the two diatomic molecules that contain no hydrogen and whose two atoms share a coordinate, namely ClF and IBr. Set B8 is an eight-molecule H/C/N/O/F panel whose set-level charge errors are reported in Table~\ref{tab:charge} and whose per-molecule hydroxyl and amide charges are reported in the Supporting Information text. It is distinct from the eight-molecule panel of the frozen-geometry sweep over functional, basis set and partition reference reported in Table~S5, which is HF, H\(_2\)O, NH\(_3\), CH\(_4\), H\(_2\)CO, CH\(_3\)OH, CO\(_2\), and SO\(_2\).

\subsection{Scoring rules}
\label{sec:verdicts}

The reference charges are the stockholder charges of Hirshfeld \cite{hirshfeld1977}. They were chosen because the stockholder partition is determined by the electron density itself rather than by the basis in which the density is expanded; Mulliken charges obtained from the same wavefunctions provide a second, deliberately different partition, and we use it in Sec.~\ref{sec:e1} to test whether our conclusions depend on that choice. Since no partition of a molecular density into atomic charges is unique, no charge comparison below should be read as measuring an observable; the charge comparisons measure agreement with one fixed and clearly stated convention. Dipole moments, unlike partial charges, are observables. The primary dipole error is nonetheless taken against the same-geometry density-functional reference, while the experimental compilation provides the secondary summary and the HX ordering check of rule~(2).

The primary measure of charge error is the mean absolute deviation of the equalization charges from the reference charges, averaged first over the atoms within a molecule and then over the molecules of a set; we refer to this two-stage average simply as the charge error. Unless stated otherwise it is evaluated on the twenty polyatomic molecules of sets B2 and B3, a combination written B2+B3 in the tables and figures that follow. Set B1 is excluded from that comparison because three of its fourteen members are homonuclear, so that every table returns zero charge there by symmetry, while the remainder mixes alkali halides with an open-shell radical; sets B4 to B6 consist of intermolecular complexes, reactivity fragments and isolated ions and are reported separately. Results for all sets appear in Table~\ref{tab:charge} and Fig.~\ref{fig:charge}. Alongside the two-stage average we report an atom-weighted mean absolute error and a root-mean-square error (RMSE) over the same 117 atoms, together with the corresponding errors against Mulliken charges. For dipole moments the primary measure is the mean absolute error of the magnitude, taken against the dipole moment computed by density-functional theory at the same geometry. The experimental compilation is retained only as a secondary summary, because its entries mix equilibrium and vibrationally averaged conventions. We also report the length of the vector difference between the two dipole moments, the cosine of the angle between them, and, for diatomic molecules, the signed projection on the bond axis. The rules that follow are the criteria by which the tables are judged.

Comparisons with reference data:
\begin{enumerate}[label=(\arabic*),leftmargin=1.6em]
\item Charges against the Hirshfeld reference on B2+B3, compared also with the trivial prediction that every atom carries zero charge, which we call the null model.
\item Dipoles. The B2+B3 dipole norm MAE against DFT is reported descriptively; the pass condition is the hydrogen halide (HX) series descending F \(>\) Cl \(>\) Br \(>\) I with the density-functional signed polarity.
\item Isolated-fragment ranking under Eq.~\eqref{eq:mismatch} against the pairing of BF\(_3\) with NH\(_3\) and BH\(_3\) with PH\(_3\) \cite{pearson1968hsab}. The prespecified form of this comparison asked instead whether L reproduces the P ranking; both forms are reported in Sec.~\ref{sec:e4}.
\end{enumerate}

Criteria (1)--(3) compare a table with reference data; criteria (A)--(C) below compare tables with one another. Throughout this paper the term \emph{scoring rule} refers to one of these six criteria and to nothing else. Criteria (1)--(3) are ordinal rather than tolerance-based, and their pass conditions are: (1) a table passes if its B2+B3 charge MAE is smaller than that of the null model defined in Sec.~\ref{sec:posthoc} \(q_i=0\); (2) a table passes if its HX dipole norms descend F \(>\) Cl \(>\) Br \(>\) I with the DFT signed polarity; (3) a table passes if it reproduces both conventional preferences. Criteria (A) and (B) carry the numerical tolerances stated below.

Comparisons among tables:
\begin{enumerate}[label=(\Alph*),leftmargin=1.6em]
\item P/L charge criterion. The two tables are judged equivalent on charges if \(\mathrm{MAE}(L)/\mathrm{MAE}(P)\) on B2+B3 lies in \([0.8,1.2]\). A ratio below \(0.8\) (L closer) or above \(1.2\) (L farther) rejects that criterion; the direction is reported separately.
\item Matched-fit comparison. Identically fitted \(\cosh\) and \(\rho^2\) meet a 5\% agreement rule if \(|\mathrm{MAE}(\mathrm{L}_{\mathrm{fit}})-\mathrm{MAE}(Q)|/\mathrm{MAE}(P)<0.05\) for both charges and dipoles versus DFT. Published-scale L versus Q is also reported. The prespecified L/Q form gives \(0.114\) and fails the criterion. \(\mathrm{L}_{\mathrm{fit}}\) was introduced subsequently to isolate kernel shape from scale provenance; its \(\mathrm{L}_{\mathrm{fit}}/Q\) value of \(0.0011\) is therefore reported as a post hoc diagnostic rather than as a prespecified pass. The two forms answer different questions: the first tests the published table against the control, whereas the second tests the kernel shapes after matching their atomic fitting protocol.
\item Shared-\(\rho\) dipoles. Experimental HX dipoles fall F \(>\) Cl \(>\) Br \(>\) I, whereas \(\chi_L(\mathrm F)=\chi_L(\mathrm{Cl})\). ClF and IBr test the same degeneracy without hydrogen.
\end{enumerate}

The sweep over the density-functional recipe and the additional set B8 serve as robustness checks on~(1) and~(B). They are not additional pass/fail criteria. Two properties of these rules affect their interpretation. Rule~(2) combines an ordering test with a sign test, so a table may fail it either for reversing every bond in the series or for placing one halogen out of sequence, and the outcome carries no record of which. Rule~(3) uses an empirical regularity as its criterion, and whether the reference calculation reproduces that regularity is a separate question from whether an atomic table does; we report the answer for the reference alongside the answer for each table (Table~\ref{tab:verdicts}) so that the weight of the rule can be judged. Neither rule is rewritten here, but two of them are not in their prespecified form: rule~(3) originally asked whether the geometric table reproduces the spectroscopic ranking rather than the conventional pairing, and rule~(B) originally compared the published geometric table with the control rather than the refitted table. Both original forms are reported alongside the revised ones, in Secs.~\ref{sec:e4} and~\ref{sec:e2q} respectively. The same status applies to the conventional-pairing form of rule~(3) as to the \(\mathrm{L}_{\mathrm{fit}}\)/Q form of rule~(B): forms introduced after the outcomes were known are treated as diagnostics rather than as passed prespecified tests.

\subsection{Null model and charge-placement measures}
\label{sec:posthoc}

The null model \(q_i=0\) has two-stage MAE equal to the mean over molecules of the within-molecule mean \(|q_i^{\mathrm{Hirsh}}|\). Where the charge is placed is scored separately from how large it is, by the following measures: the Pearson and Spearman correlations between the EEM and Hirshfeld charge vectors within each molecule (molecules with at least three atoms and standard deviation of both vectors above \(10^{-6}\,e\)), the fraction of non-hydrogen atoms with \(|q_i^{\mathrm{Hirsh}}|\ge0.02\,e\) whose EEM sign matches Hirshfeld, and the balanced sign rate \(\tfrac12(\mathrm{TPR}+\mathrm{TNR})\), the mean of the true-positive rate (TPR) and the true-negative rate (TNR), on those atoms. Hydrogen is omitted from the sign count. Cosine of the angle between EEM and DFT dipole vectors is defined only when both norms exceed \(10^{-6}\,\mathrm{D}\); undefined cases are excluded and the sample size is reported. A negative cosine means an angle greater than \(90^\circ\), not exact collinear opposition. Charge errors on B2+B3 are also split by hydrogen versus geometric atoms and by hydrogen-containing versus hydrogen-free molecules. Nearest-neighbor bond classes (O--H, N--H, C--H, S--H, P--H, H--X, C--X) are tallied on B1--B3.

The Hirshfeld charges, placed as point charges at the same nuclei, supply a reference monopole baseline for the dipole comparison. Bootstrap, Wilcoxon and constrained-Hessian numerics are recorded in the Supporting Information, including the smallest eigenvalue of the constrained Hessian under each of the four tables and the one molecule, KCl, whose molecular response slope changes sign; the no-Coulomb limit is reported in Sec.~\ref{sec:theory} instead, because removing the coupling changes the ranking of the tables and is therefore not a side diagnostic. The main-text claim does not rest on those diagnostics. No conclusion below is obtained by counting the six criteria as independent pieces of evidence; Sec.~\ref{sec:discussion} separates the criteria that discriminate among the tables from those that do not.

\subsection{Scope and limitations}
\label{sec:scope}

Several features bound the interpretation of the comparisons that follow. First, a frozen table is not a parameter-free table: the geometric construction carries atomic-scale constants fitted to spectroscopic data, so the comparison concerns alternative ways of assigning atomic parameters rather than a parameter-free model set against a fitted one. Second, the molecule list was assembled for this study by the group that proposed the coordinate, and no held-out subset was defined. Positive comparisons are therefore sensitive to the composition of the list; that sensitivity is realized here by the P/L charge ordering, which reverses on the separate eight-molecule panel of Table~S5, whereas the shared-coordinate identity of Sec.~\ref{sec:eem} holds for any list whatsoever. This design limitation bears most directly on favorable numerical comparisons; the exact shared-coordinate identity of Sec.~\ref{sec:eem} is independent of molecule-list composition.

Atomic charges introduce a separate limitation, because their values depend on the partitioning convention. Hirshfeld charges provide the primary charge reference and Mulliken charges a deliberately different cross-check, so no directional conclusion is drawn from the P/L charge comparison when the two treatments do not support the same operational verdict. Dipole moments are treated separately, as observables.

The numerical equivalence tolerances in criteria~(A) and~(B) are conventions of this study rather than external standards. Where an interval or a paired test conflicts with a threshold, the statistical interval controls the interpretation.

Finally, the Br/I all-electron/ECP asymmetry and the mixed-basis finite-energy hardness reference are retained limitations of the electronic-structure protocol, as described in Sec.~\ref{sec:dft}. They can move the numerical references against which a table is scored, but neither determines the exact shared-coordinate dipole result.

\section{Results}
\label{sec:results}

Geometry optimizations and property evaluations were completed for every species of sets B1 to B7 in Table~\ref{tab:mols} apart from the bare proton, which has no electrons; this gives fifty-two molecules and ions, including the two hydrogen-free diatomics of set B7, and a further eight organic molecules in set B8. The equalization equations were solved for all four tables at each optimized geometry, and every linear system proved well conditioned.

\subsection{Atomic inputs before equalization}

\begin{figure}[!ht]
\centering
\includegraphics[width=0.92\textwidth]{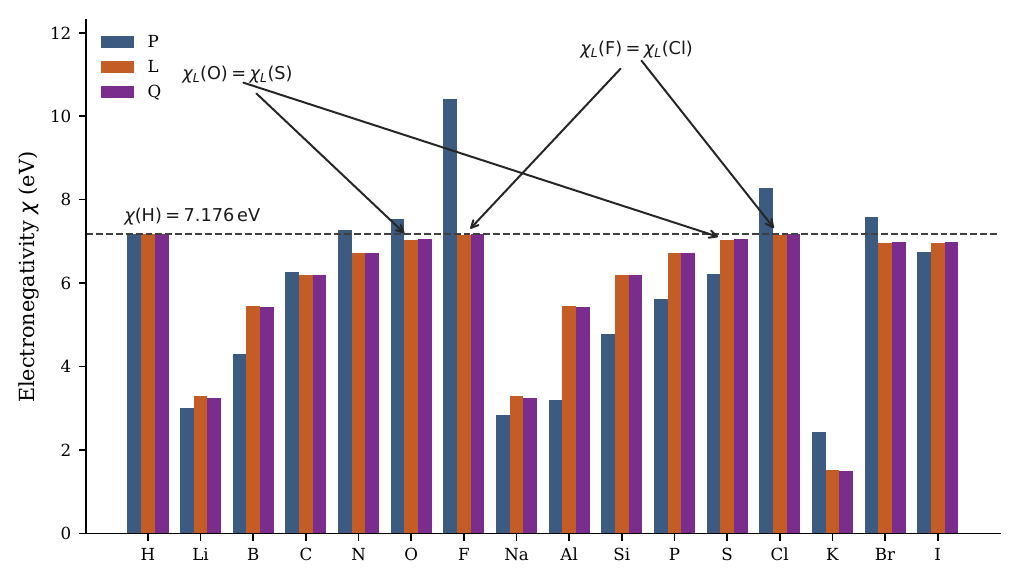}
\caption{Atomic electronegativities of hydrogen, which keeps its spectroscopic value, and of the atoms the geometric construction describes. L and Q compress the main-group values below \(\chi(\mathrm{H})=7.176\,\mathrm{eV}\) (dashed) and assign one electronegativity to every pair of atoms plotted here that shares \(d/L_p\): Li/Na, B/Al, C/Si, N/P, O/S, F/Cl, and Br/I. The hardnesses stay distinct within each pair, because \(C_\eta\) is resolved by period. Those two features of the table, not the molecular solver, produce the additional O--H, N--H, and hydrogen-halide inversions (C--H and S--H already have that ordering under P) and the HF/HCl dipole collapse.}
\label{fig:atomic}
\end{figure}

Figure~\ref{fig:atomic} plots the entries of Table~\ref{tab:atomic} and shows the two properties of the atomic table established in Sec.~\ref{sec:rho} (the isolated position of hydrogen and the seven degenerate pairs) as they stand before any molecule is considered.

\FloatBarrier

\subsection{Dipole moments of molecules whose atoms share a coordinate}
\label{sec:hx}

\begin{figure}[!ht]
\centering
\includegraphics[width=0.92\textwidth]{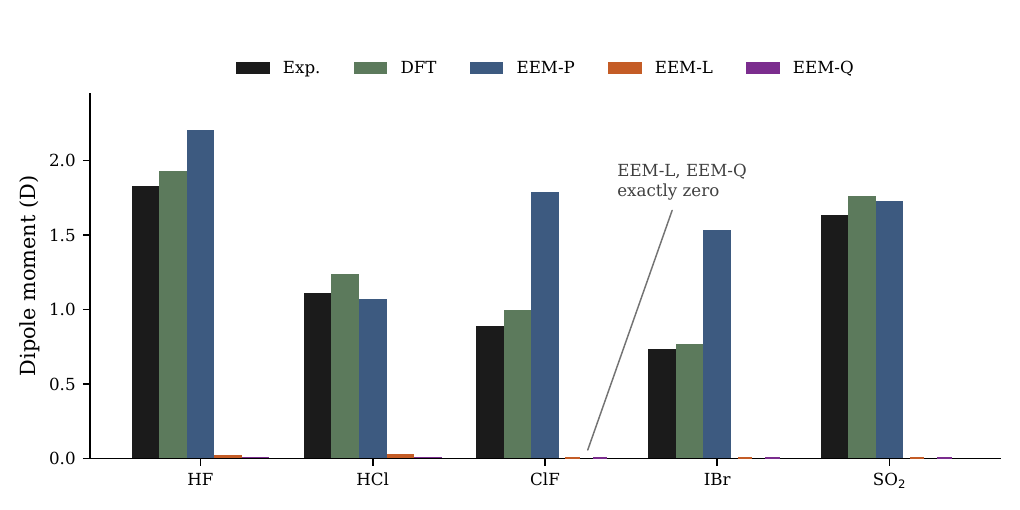}
\caption{Dipole comparison highlighting the shared-coordinate collapse. Experiment and density-functional theory remain finite, while the geometric tables give exactly zero for ClF, IBr and SO\(_2\) (Tables~\ref{tab:hx} and~\ref{tab:b7}). \(\mathrm{L}_{\mathrm{fit}}\), omitted from the plot for clarity, gives the same zero values.}
\label{fig:carry}
\end{figure}

\begin{figure}[!ht]
\centering
\includegraphics[width=0.92\textwidth]{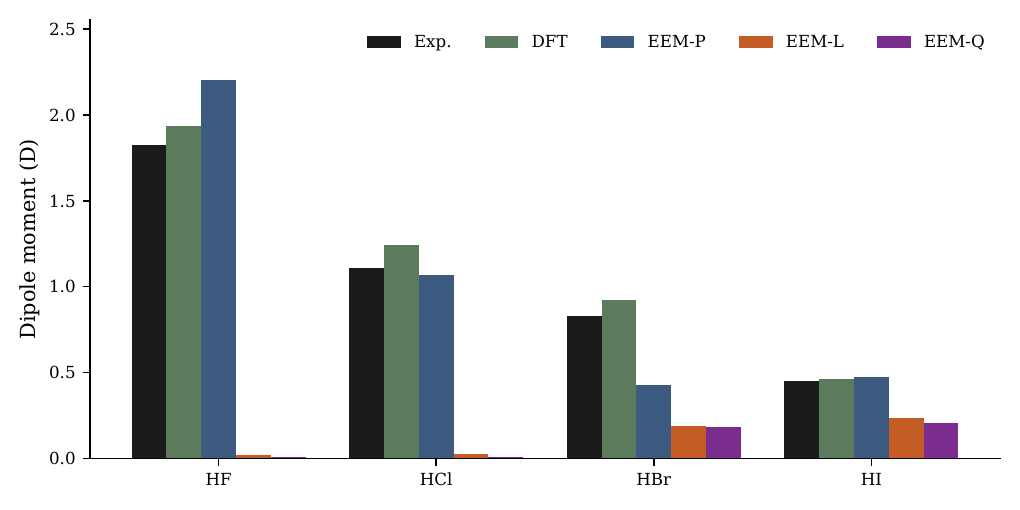}
\caption{Hydrogen halide dipole norms. Experiment and DFT descend F \(>\) Cl \(>\) Br \(>\) I. EEM-P retains a large HF/HCl split. EEM-L and EEM-Q collapse HF and HCl because \(\chi(\mathrm{F})=\chi(\mathrm{Cl})\) in both geometric tables.}
\label{fig:hx}
\end{figure}

The hydrogen halides provide the simplest test of the shared coordinate, because the measured dipole moments of the four molecules decrease monotonically from HF to HI while two of the four halogens are assigned identical electronegativities. Figure~\ref{fig:carry} gathers the five molecules in which the shared-coordinate effect can be seen directly. The hydrogen-halide magnitudes are listed in Table~\ref{tab:hx} and shown separately in Fig.~\ref{fig:hx}, while the corresponding hydrogen-free pair appears in Table~\ref{tab:b7}. Density-functional theory reproduces the experimental ordering. The spectroscopic table retains a large separation between HF and HCl and a decreasing sequence from fluorine to bromine, with HI marginally above HBr. The geometric table does not: because fluorine and chlorine are given the same electronegativity, the HF and HCl magnitudes collapse onto one another, exactly as the two-atom solution of Eq.~\eqref{eq:qh-hf} requires, and what separation remains comes from the differing halogen hardnesses and bond lengths. The refitted and quadratic tables behave in the same way. Moreover the signed projections on the bond axis are reversed relative to density-functional theory for all four hydrogen halides under those three tables, so the collapse is an inversion of polarity and not merely a small dipole moment of the correct sign. The degeneracy identified for free atoms is therefore inherited by a molecular observable. Scoring rule~(2) as stated in Sec.~\ref{sec:verdicts} is nonetheless failed by all four tables and not only by the three geometric ones. Under the spectroscopic table the sequence breaks at its last step, HI (\(0.476\,\mathrm{D}\)) standing above HBr (\(0.425\,\mathrm{D}\)), and the signed projection for HI is \(+0.476\,\mathrm{D}\) where density-functional theory gives \(-0.462\,\mathrm{D}\). The two failures are of quite different size, one halogen out of order against four bonds of reversed polarity, but the rule admits no partial credit and we do not claim a pass for the spectroscopic table here.


\FloatBarrier

Alkali halides show the complementary pattern. DFT tracks experiment (LiF \(6.13\) versus \(6.33\,\mathrm{D}\); KCl \(10.15\) versus \(10.27\,\mathrm{D}\)). EEM-P overshoots (LiF \(9.87\,\mathrm{D}\); KCl \(15.54\,\mathrm{D}\)). EEM-L is closer than EEM-P on three of the four alkali halides, on LiF (\(4.66\,\mathrm{D}\)), on LiCl (\(8.03\,\mathrm{D}\) against \(12.53\,\mathrm{D}\) for P and \(6.93\,\mathrm{D}\) for DFT) and NaCl (\(9.28\) versus experiment \(9.00\,\mathrm{D}\)) and worse on KCl (\(17.41\,\mathrm{D}\)), where \(\eta_L(\mathrm{K})\) is small enough that \(\Kmol\) changes sign (Sec.~\ref{sec:etamol}). The two matched tables are worse still on this set, and by a wide margin: on potassium chloride \(\mathrm{L}_{\mathrm{fit}}\) and Q give \(23.93\) and \(24.08\,\mathrm{D}\) against \(10.15\,\mathrm{D}\) from density-functional theory. The agreement of the two kernels established in Sec.~\ref{sec:e2q} is an agreement with each other on the twenty polyatomic molecules and not a claim that either reproduces the ionic diatomics. Per-molecule alkali-halide dipoles are in Table~S3; Table~S1 holds the organohalides only. These four alkali halides are diatomic and lie outside sets B2 and B3, so they enter the paper only as a signed check on the direction of the dipole and contribute nothing to the dipole error of Table~\ref{tab:place}.

\begin{table}[!ht]
\centering
\caption{HX dipole norms in debye.}
\label{tab:hx}
\small
\begin{tabular}{lrrrrrr}
\toprule
Molecule & Exp. & DFT & EEM-P & EEM-L & EEM-\(\mathrm{L}_{\mathrm{fit}}\) & EEM-Q \\
\midrule
HF & 1.826 & 1.932 & 2.203 & 0.020 & 0.019 & 0.006 \\
HCl & 1.109 & 1.240 & 1.070 & 0.028 & 0.026 & 0.009 \\
HBr & 0.827 & 0.922 & 0.425 & 0.189 & 0.198 & 0.184 \\
HI & 0.448 & 0.462 & 0.476 & 0.233 & 0.225 & 0.208 \\
\bottomrule
\end{tabular}
\tabnotes{Experimental values from CCCBDB \cite{cccbdb}. DFT is B3LYP/def2-TZVP at the optimized geometry. EEM dipoles use DFT frames and Ohno charges.}
\end{table}

\FloatBarrier

\begin{figure}[!ht]
\centering
\includegraphics[width=0.92\textwidth]{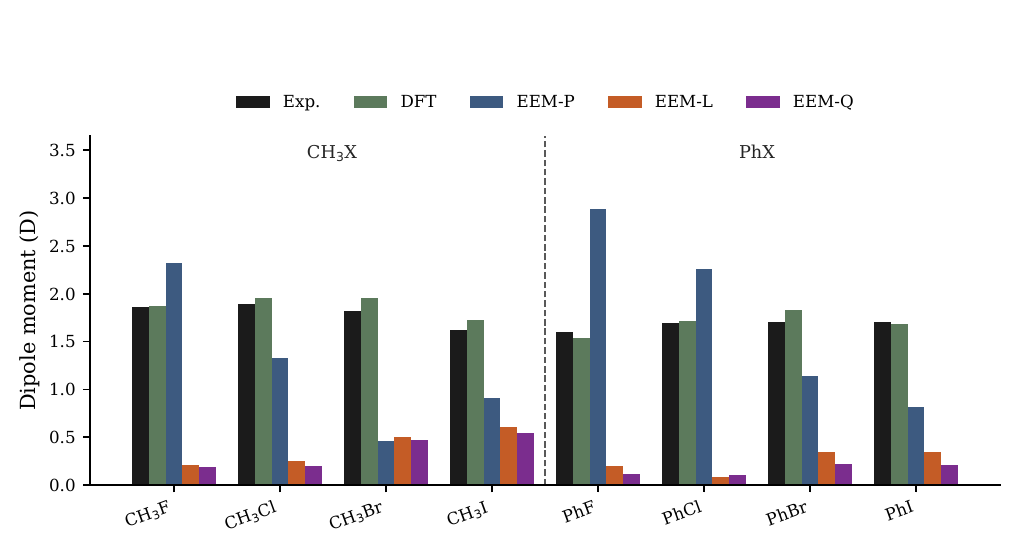}
\caption{Methyl and phenyl halide dipole norms. The dashed line separates the CH\(_3\)X series (left) from the PhX series (right). Experiment and DFT remain near \(1.5\)--\(2.0\,\mathrm{D}\) in both series. The geometric tables stay small and show no consistent ordering of fluorine against chlorine, the polyatomic counterpart of the HF/HCl collapse in Fig.~\ref{fig:hx}.}
\label{fig:ch3x}
\end{figure}

Methyl and phenyl halides are the polyatomic analogues in which the same degeneracy can appear (Fig.~\ref{fig:ch3x}; numerical values in Table~S1). Experiment on CH\(_3\)X does not follow F \(>\) Cl (\(1.858\), \(1.892\), \(1.822\), \(1.62\,\mathrm{D}\)). DFT instead gives Br \(\gtrsim\) Cl \(>\) F \(>\) I, with Br and Cl differing by only \(0.002\,\mathrm{D}\). EEM-L dipoles on the eight organohalides are \(0.08\)--\(0.60\,\mathrm{D}\) and do not recover a consistent F/Cl ordering: the geometric bars remain small across both series, while experiment and DFT stay near \(1.5\)--\(2.0\,\mathrm{D}\). EEM-P overshoots PhF (\(2.89\,\mathrm{D}\) versus experiment \(1.60\,\mathrm{D}\)) and is not an independent down-group test because \(\chi_P(\mathrm{F})\neq\chi_P(\mathrm{Cl})\) already contains the spectroscopic ordering. Chlorine monofluoride and iodine monobromide contain no hydrogen, and each is built entirely from a single degenerate pair. Equal geometric electronegativity forces the zero-charge solution of Eq.~\eqref{eq:eem}; Table~\ref{tab:b7} is the numerical instance of the collapse already shown in Fig.~\ref{fig:carry}. ClF, IBr and SO\(_2\) are chemical instances of the single identity established in Sec.~\ref{sec:eem}, and are therefore not treated as statistically independent failures.

\begin{table}[!ht]
\centering
\caption{Dipole norms of ClF and IBr in debye.}
\label{tab:b7}
\small
\begin{tabular}{lrrrrrr}
\toprule
Molecule & Exp. & DFT & EEM-P & EEM-L & EEM-\(\mathrm{L}_{\mathrm{fit}}\) & EEM-Q \\
\midrule
ClF & 0.888 & 0.998 & 1.790 & 0.000 & 0.000 & 0.000 \\
IBr & 0.737 & 0.768 & 1.534 & 0.000 & 0.000 & 0.000 \\
\bottomrule
\end{tabular}
\tabnotes{Experimental values from CCCBDB \cite{cccbdb}. Because the two atoms of each molecule are assigned the same geometric electronegativity, the equalization equations are satisfied by zero charge on both atoms, for the published, the refitted and the quadratic table alike.}
\end{table}

\FloatBarrier

\subsection{Identically fitted hyperbolic and quadratic kernels}
\label{sec:e2q}
We next test whether the kernel shape has a detectable effect on the molecular predictions. The published construction uses a hyperbolic cosine of the coordinate, and the control replaces it by the square of the same coordinate; if the two agree once their energy scales are fitted by the same protocol, then the molecular data do not resolve an effect attributable to the golden-ratio kernel. On the twenty polyatomic molecules the two agree closely:
\begin{equation}
\mathrm{MAE}(\mathrm{L}_{\mathrm{fit}})=0.20089\,e,\qquad
\mathrm{MAE}(Q)=0.20114\,e,
\label{eq:lfitq}
\end{equation}
so \(|\mathrm{MAE}(\mathrm{L}_{\mathrm{fit}})-\mathrm{MAE}(Q)|/\mathrm{MAE}(P)=0.0011\). The comparison in Eq.~\eqref{eq:lfitq} is a difference of two averaged errors, and it is worth adding the stronger comparison the same data allow, because a difference of averages can be small even when two models disagree atom by atom. Comparing the two charge predictions directly, atom against atom on the same 117 atoms, gives a mean absolute difference of \(0.0022\,e\), which is \(0.0098\) of the spectroscopic error. Published-scale L against Q gives \(0.0348\,e\) on the same pointwise measure, that is \(0.155\). The verdict is therefore the same under either measure, matched kernels inside the tolerance by a factor of five and published-scale L outside it by a factor of three, and the pointwise figures are the ones that cannot be attributed to cancellation. The corresponding dipole MAEs versus DFT are \(1.361\,\mathrm{D}\) and \(1.370\,\mathrm{D}\) (normalized difference \(0.0089\)). Published-scale L versus Q, by contrast, gives \(|\mathrm{MAE}(L)-\mathrm{MAE}(Q)|/\mathrm{MAE}(P)=0.114\). L and Q do not share a fitting protocol: L uses the published period-2--4 hardness scales, while Q (and \(\mathrm{L}_{\mathrm{fit}}\)) are through-origin fits on the atoms of Table~\ref{tab:atomic}. Once those scales are equalized, the hyperbolic advantage on charges disappears. The published-scale L/Q gap is therefore a scale-provenance effect.

\begin{figure}[!ht]
\centering
\includegraphics[width=0.40\textwidth]{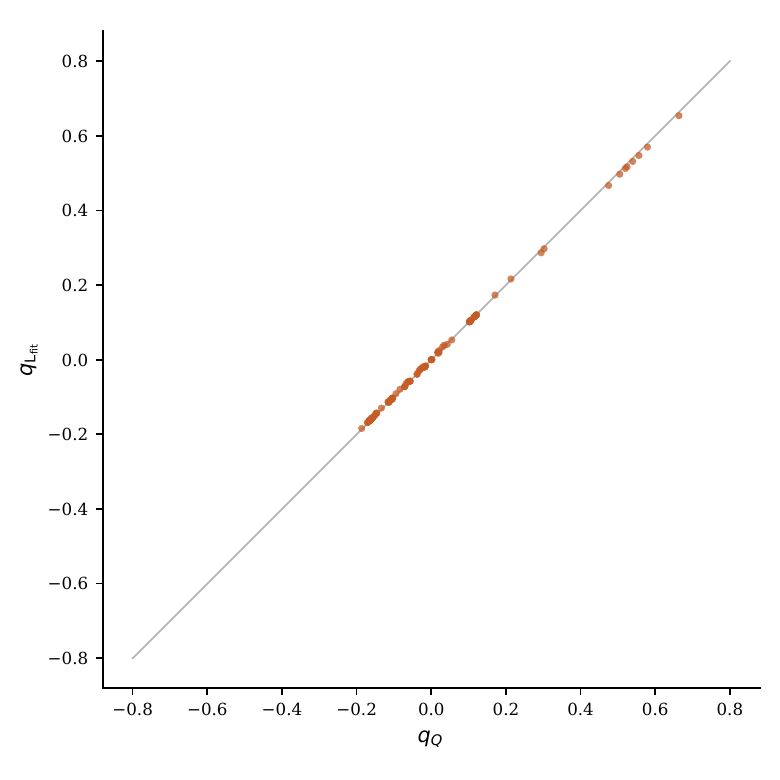}
\caption{Atomic charges on B2+B3 from identically fitted \(\cosh\) (\(\mathrm{L}_{\mathrm{fit}}\)) and \(\rho^2\) (Q) kernels. The two predictions lie on the diagonal (MAE \(0.0022\,e\)).}
\label{fig:lfitq}
\end{figure}

The agreement is not only of average errors. On the 117 atoms of B2+B3 the \(\mathrm{L}_{\mathrm{fit}}\) and Q charges differ by MAE \(0.0022\,e\), RMS \(0.0031\,e\), and maximum \(0.0099\,e\); their Pearson and Spearman correlations are \(0.99996\) and \(0.9995\). Molecular dipole norms differ by \(0.014\,\mathrm{D}\) on average (maximum \(0.043\,\mathrm{D}\)), and the mean cosine between the two dipole vectors is \(0.99998\) (\(n=17\); CH\(_4\), CO\(_2\), and SO\(_2\) are excluded because at least one norm vanishes). The fitted atomic tables themselves are already close: the geometric \(\chi\) and \(\eta\) differ by \(0.018\,\mathrm{eV}\) and \(0.010\,\mathrm{eV}\) on average. After identical atomic fitting the two kernels therefore propagate into almost the same molecular predictions, not merely the same MAEs. Q retains the same \(\rho\), including the F/Cl and O/S degeneracies, and both kernels remain close on dipoles either way. These molecular comparisons do not distinguish the \(\phig\)-based hyperbolic kernel from the matched quadratic control, and therefore supply no chemical support for the hyperbolic form as a molecular descriptor. Figure~\ref{fig:lfitq} shows the same result atom by atom: over all of B2+B3 the charges predicted by the two identically fitted kernels fall on the diagonal, with a mean absolute difference of \(0.0022\,e\), so the agreement is not confined to the set-level averages.


\FloatBarrier

The comparison of two kernels can be extended to a continuum of them without any new density-functional calculation. For each base \(b\) we replaced the kernel by \(J_b(\rho)=\cosh(\rho\ln b)-1\), refitted its energy scales by the same through-origin protocol, and re-solved the equalization problem on B2+B3; the choice \(b=\phig\) reproduces \(\mathrm{L}_{\mathrm{fit}}\) exactly. Across \(b=1.05\) to \(10\) the two-stage charge error spans only \(0.2005\) to \(0.2062\,e\). The golden ratio gives \(0.2009\,e\), and the shallow minimum falls near \(b\approx2.7\) to \(3.2\) rather than at \(\phig\). These molecules therefore do not identify a preferred base anywhere in that range, which is a stronger statement than the failure to separate two particular kernels. Individual values are listed in the Supporting Information.

\FloatBarrier

\subsection{Partial charges and dipoles}
\label{sec:e1}

We next compare the equalization charges with the density-based reference, and begin by distinguishing the two halves of that comparison, which are of very unequal strength. Comparing the geometric table with the spectroscopic one, criterion~(A), is the weakest comparison in the paper. Comparing every table with the zero-charge model, rule~(1), is among the strongest. The reasons that follow apply to the first and not to the second. Criterion~(A) is the only comparison whose outcome changes when the partition is changed, and three of the quantities being compared are not independent of one another: hydrogen retains its spectroscopic electronegativity in every table and lies above all geometric values, silver is copied from the spectroscopic table into every other, and the period-5 hardness scale was fixed by iodine alone, so that iodine reproduces its spectroscopic hardness by construction. This last point matters because iodine then appears in HI, CH\(_3\)I, PhI, IBr and set B6 as though it were testing the geometric hardness scale: the through-origin fit of Eq.~\eqref{eq:through-origin} over a single atom returns \(\eta_L(\mathrm{I})=\eta_P(\mathrm{I})=3.6961\,\mathrm{eV}\) identically, and silver is copied across in the same way at \(3.1359\,\mathrm{eV}\). Every comparison involving iodine therefore tests the geometric electronegativity and the spectroscopic hardness in combination, never the geometric hardness on its own. Conclusions drawn below from iodine-containing species rest on the electronegativity degeneracy Br/I, which is unaffected, and not on the period-5 hardness constant.

\begin{figure}[!ht]
\centering
\includegraphics[width=0.92\textwidth]{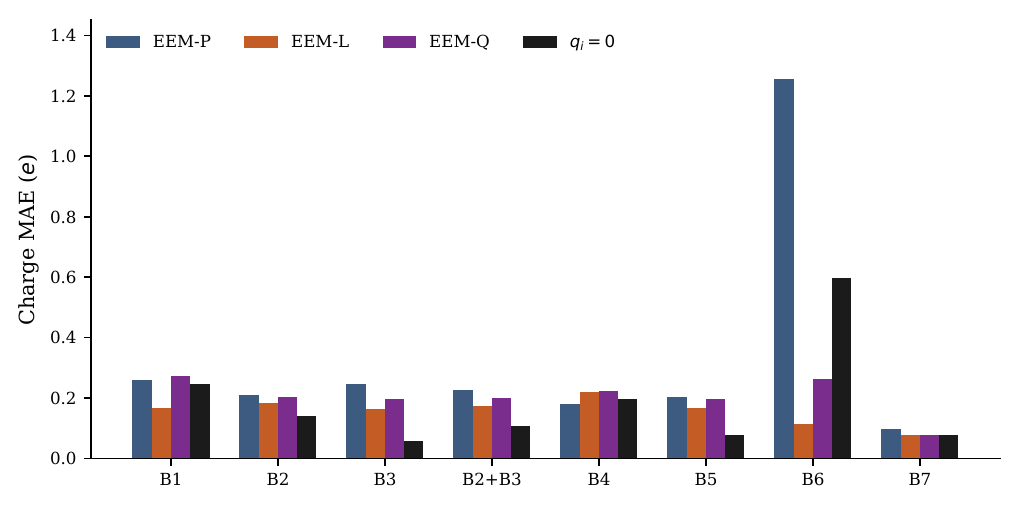}
\caption{Two-stage charge MAE versus Hirshfeld by molecule set (B1--B7). The null model \(q_i=0\) is the closest column on B2, B3, B2+B3, and B5, closer than P, L, \(\mathrm{L}_{\mathrm{fit}}\), and Q. L is the closest EEM table except on B4. The B6 spike is Ohno-coupled overpolarization of table~P on BF\(_3\), BH\(_3\), and AlCl\(_3\). B8 is in Table~\ref{tab:charge} but is not plotted.}
\label{fig:charge}
\end{figure}

Table~\ref{tab:charge} and Fig.~\ref{fig:charge} report the two-stage charge MAE versus Hirshfeld.

\begin{table}[!ht]
\centering
\caption{Mean per-molecule MAE of EEM charges versus DFT Hirshfeld charges, in \(e\).}
\label{tab:charge}
\small
\begin{tabular}{lrrrrrr}
\toprule
Set & \(n\) & MAE-P & MAE-L & MAE-\(\mathrm{L}_{\mathrm{fit}}\) & MAE-Q & \(q_i=0\) \\
\midrule
B1 & 14 & 0.259 & 0.166 & 0.266 & 0.272 & 0.245 \\
B2 & 12 & 0.210 & 0.184 & 0.205 & 0.204 & 0.139 \\
B3 & 8 & 0.248 & 0.163 & 0.195 & 0.196 & 0.057 \\
B2+B3 & 20 & 0.225 & 0.175 & 0.201 & 0.201 & 0.106 \\
B4 & 4 & 0.180 & 0.220 & 0.226 & 0.223 & 0.197 \\
B5 & 6 & 0.202 & 0.168 & 0.195 & 0.196 & 0.079 \\
B6 & 6 & 1.255 & 0.113 & 0.249 & 0.262 & 0.598 \\
B7 & 2 & 0.099 & 0.078 & 0.078 & 0.078 & 0.078 \\
B8 & 8 & 0.246 & 0.182 & 0.218 & 0.219 & 0.104 \\
\bottomrule
\end{tabular}
\tabnotes{B2+B3 is the set for the P/L charge comparison. B5 here is the six organics not already in B2 (H\(_2\)CO, CO\(_2\) counted under B2). B7 is ClF and IBr. B8 is the eight-molecule H/C/N/O/F panel listed in Table~\ref{tab:mols}; it is not the recipe-sweep panel of Table~S5 and is not used for the scoring rules. \(\mathrm{L}_{\mathrm{fit}}\) uses the hyperbolic kernels with the same through-origin scales as Q. The last column is the null model \(q_i=0\). Per-molecule values for B1--B7 are in Table~S2.}
\end{table}

\FloatBarrier

On B2+B3,
\begin{equation}
\mathrm{MAE}(P)=0.22519\,e,\quad
\mathrm{MAE}(L)=0.17549\,e.
\end{equation}
The matched \(\mathrm{L}_{\mathrm{fit}}\) and \(Q\) pair is Sec.~\ref{sec:e2q}. Those five-digit values are the two-stage means; Table~\ref{tab:charge} prints the same quantities to three digits (\(0.225\), \(0.175\)), and \(0.17549/0.22519=0.7793\).
The ratio \(\mathrm{MAE}(L)/\mathrm{MAE}(P)=0.779\) lies below the \(0.8\) equivalence-interval bound. That rejection is not robust.


\FloatBarrier

A molecule-level bootstrap \(95\%\) interval for the ratio is \([0.618,1.034]\) (\texttt{data/verdicts.json}). The interval contains the equivalence-interval edge \(0.8\) and it contains unity, and on resampling the twenty molecules the ratio falls inside the equivalence interval \([0.8,1.2]\) in \(40\%\) of draws. A point estimate whose interval covers the decision threshold cannot be reported as a clean categorical outcome, and Table~\ref{tab:verdicts} therefore records criterion~(A) as unresolved rather than rejected. The same caution applies to the direction of the comparison: paired over the twenty molecules the spectroscopic table is worse than the geometric one by \(0.050\,e\) on average, but the \(95\%\) interval on that difference is \([-0.006,+0.100]\,e\) and contains zero (Wilcoxon \(p=0.048\), the geometric table closer on 16 of 20 molecules). The geometric table is therefore not shown to be better than the spectroscopic one on this set; it is not shown to be worse either. Against Mulliken charges of the same densities the ratio moves from \(0.779\) to \(0.868\). Bootstrap and Wilcoxon details are in the Supporting Information. For comparison, an equalization model whose parameters have been fitted to quantum-chemical charges reaches an error of \(0.02\)--\(0.06\,e\) on comparable sets \cite{bultinck2002eem}; none of the frozen tables approaches that accuracy, which is only to be expected, since none of them was fitted to a molecule. Atom-weighted MAE on the same 117 atoms is \(0.213\,e\) (P), \(0.163\,e\) (L), \(0.189\,e\) (Q); RMSE is \(0.286\), \(0.189\), and \(0.231\,e\), respectively.

The set-level pattern in Fig.~\ref{fig:charge} and Table~\ref{tab:charge} is that the null model \(q_i=0\) is the closest column on B2, B3, their union, B5, and B8, while L is the closest EEM table on every set except the four dimers of B4. On B6, which excludes H\(^+\), EEM-P places \(q_{\mathrm{B}}\approx+6.1\) on BF\(_3\) (Hirshfeld \(+0.46\)) and \(+4.9\) and \(+5.2\) on the central atoms of BH\(_3\) and AlCl\(_3\). The first of those values lies outside the physically available range: boron has five electrons, so a charge of \(+6.1\) implies transferring more charge than the atom carries, and \(+4.9\) in borane amounts to complete ionization. This is a breakdown of the quadratic energy expression rather than a large error within it, since Eq.~\eqref{eq:energy} contains nothing that limits \(q_i\) to the available electron count; it is the reason set B6 is reported for completeness and excluded from every scoring rule. EEM-L gives \(q_{\mathrm{B}}=+0.71\) on BF\(_3\) and \(+1.04\) on BH\(_3\), with the Hirshfeld sign and much smaller magnitude. Thus the B6 MAE is \(1.26\,e\) for P and \(0.11\,e\) for L, from Ohno-coupled overpolarization of the P table on BF\(_3\), BH\(_3\), and AlCl\(_3\). That set mean should not be quoted without its composition, and we therefore do not interpret it as a set-level result. Three of the six members of B6 are monatomic ions, for which a single atom carries the whole net charge and every table is exact by construction, giving an error of identically zero; the other three are the trigonal molecules, with P errors of \(2.33\), \(2.83\) and \(2.37\,e\). The mean of \(1.26\,e\) is an average over two populations with nothing in common and describes neither. The overpolarization claim refers to the three trigonal molecules only, where it is plain in the per-molecule charges, and no test on six points is offered in support of it. The same caution applies to the other small sets: on B4 (\(n=4\)) the difference between the two tables is not significant (Wilcoxon \(p=0.88\)), and B7 (\(n=2\)) admits no interval at all, so the statements made about those sets are descriptions of the individual molecules and not inferences about a population. On B7 the geometric columns coincide with the null model because the EEM charges vanish. Set B4 is the only set on which the spectroscopic table is the closer of the two, at \(0.180\,e\) against \(0.220\,e\), a difference that four molecules cannot establish (Wilcoxon \(p=0.88\); the per-molecule spectroscopic errors run \(0.09\) to \(0.36\,e\)), and it is also the set on which no table is of any use: the four hydrogen-bonded dimers have density-functional dipole moments of \(3.4\) to \(4.8\,\mathrm{D}\), of which the spectroscopic table recovers at most \(3.1\,\mathrm{D}\) and the geometric table at most \(0.3\,\mathrm{D}\) (Table~S3). The charge that flows between two molecules held together by a hydrogen bond is not something a table of isolated-atom parameters is built to supply, and we therefore report B4 for completeness and draw no conclusion from the ordering of the tables on it.

The null MAE of \(0.106\,e\) on B2+B3 lies below \(0.175\,e\) for L and \(0.225\,e\) for P. Because this is the principal statistical comparison in the charge analysis, it is assessed with a paired test rather than by comparing means alone. Pairing each table against the zero-charge model molecule by molecule over the twenty molecules of B2+B3, the mean excess error is \(+0.119\,e\) for P (\(95\%\) interval \([+0.047,+0.186]\), Wilcoxon \(p=2.1\times10^{-3}\), worse on 15 of 20 molecules), \(+0.069\,e\) for L (\([+0.038,+0.096]\), \(p=7.4\times10^{-4}\), 18 of 20), \(+0.095\,e\) for \(\mathrm{L}_{\mathrm{fit}}\) (\([+0.051,+0.134]\), \(p=7.4\times10^{-4}\), 18 of 20) and \(+0.095\,e\) for Q (\([+0.050,+0.135]\), \(p=7.4\times10^{-4}\), 18 of 20). All four intervals exclude zero, and the one-sided sign tests agree (\(p\le0.021\)). Sulfur dioxide is a tie under the three geometric tables rather than a loss, because their equalization charges vanish there and reproduce the null model exactly. Unlike the P/L comparison above, this result is statistically secure. Where the reference charges are large the ordering reverses: on the alkali halides and diatomics of B1 the null MAE is \(0.245\,e\) against \(0.166\,e\) for L, and on the ionic fragments of B6 it is \(0.598\,e\) against \(0.113\,e\). This comparison concerns agreement with the Hirshfeld charge convention, and should not be read as evidence that zero charge is a better physical model of molecular electrostatics. The null model fails every polar dipole by construction; charge-partition error and dipole error answer different questions.

\begin{figure}[!ht]
\centering
\includegraphics[width=0.92\textwidth]{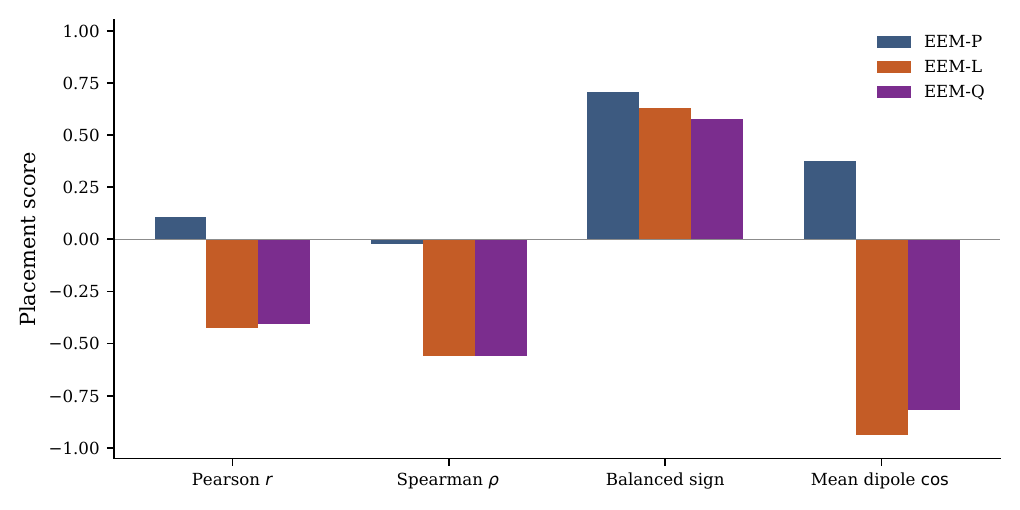}
\caption{Placement scores on B2+B3. The within-molecule Pearson and Spearman correlations of the EEM and Hirshfeld charge vectors are strongly negative under L and Q (mean \(r=-0.42\) and \(-0.40\); mean \(\rho_s=-0.56\) for both), whereas under P they are near zero (\(r=+0.11\), \(\rho_s=-0.02\)). The mean cosine of the EEM and DFT dipoles follows the same sign pattern, so the lower Hirshfeld MAE of L in Fig.~\ref{fig:charge} is a magnitude coincidence, not a better arrangement of charge.}
\label{fig:place}
\end{figure}

The primary dipole MAE on B2+B3 versus DFT is \(1.03\,\mathrm{D}\) (P), \(1.34\,\mathrm{D}\) (L), \(1.37\,\mathrm{D}\) (Q) (\(\mathrm{L}_{\mathrm{fit}}\) \(1.361\,\mathrm{D}\) against Q \(1.370\,\mathrm{D}\)). The secondary summary against the experimental compilation, promised in Sec.~\ref{sec:verdicts}, gives \(0.99\), \(1.27\) and \(1.30\,\mathrm{D}\) for the same three tables, so the ordering and the size of the gap are the same whichever reference is used and nothing in the argument turns on the choice. The vector error \(\|\boldsymbol{\mu}_{\mathrm{EEM}}-\boldsymbol{\mu}_{\mathrm{DFT}}\|\) is \(1.42\,\mathrm{D}\) (P) and \(1.81\,\mathrm{D}\) (L); the mean cosine between the EEM and DFT dipoles is \(+0.38\) for P (\(n=18\)) and \(-0.94\) for L (\(n=17\)), and the dipole cosine is negative for all \(17\) molecules with a defined cosine under L. Q and \(\mathrm{L}_{\mathrm{fit}}\) have a negative cosine on \(16\) of those \(17\). The excluded cases are the symmetry-zero DFT dipoles of CH\(_4\) and CO\(_2\) and, under L, Q, and \(\mathrm{L}_{\mathrm{fit}}\), the identically zero EEM dipole of SO\(_2\). A magnitude-only MAE therefore misses the polarity inversion that the charge analysis identifies. The Hirshfeld charges themselves, placed as point charges at the same nuclei, give a mean dipole of \(0.927\,\mathrm{D}\) against the DFT value of \(1.585\,\mathrm{D}\). Those monopoles recover \(56\%\) of the DFT dipole on the polar members of the set and miss the DFT vector by \(0.658\,\mathrm{D}\) on average, which provides a reference for the representation error carried by atom-centered Hirshfeld monopoles before the choice of atomic table enters at all. For water, DFT gives \(2.06\,\mathrm{D}\) (exp.\ \(1.855\,\mathrm{D}\)), EEM-P \(0.32\,\mathrm{D}\), and EEM-L \(0.08\,\mathrm{D}\) (Table~S3). Sulfur dioxide is the O/S analogue of F/Cl: \(\chi_L(\mathrm{O})=\chi_L(\mathrm{S})=7.036\,\mathrm{eV}\), and when all \(\chi_i^0\) in a neutral molecule coincide, \(\mathbf q=0\) satisfies Eq.~\eqref{eq:eem} together with the charge constraint, so EEM-L and EEM-Q give a vanishing dipole against experiment \(1.63\,\mathrm{D}\) (DFT \(1.76\,\mathrm{D}\); EEM-P \(1.73\,\mathrm{D}\)). Table~\ref{tab:place} and Fig.~\ref{fig:place} collect the placement scores.
\begin{table}[!ht]
\centering
\caption{Charge-placement and dipole-direction scores on B2+B3.}
\label{tab:place}
\small
\begin{tabular}{lrrrr}
\toprule
Measure & P & L & \(\mathrm{L}_{\mathrm{fit}}\) & Q \\
\midrule
Pearson \(r\) & \(+0.107\) & \(-0.423\) & \(-0.407\) & \(-0.404\) \\
Spearman \(\rho\) & \(-0.021\) & \(-0.560\) & \(-0.549\) & \(-0.559\) \\
Balanced sign & 0.71 & 0.63 & 0.58 & 0.58 \\
Dipole norm MAE vs DFT (D) & 1.03 & 1.34 & 1.36 & 1.37 \\
Dipole vector MAE vs DFT (D) & 1.42 & 1.81 & 1.80 & 1.79 \\
Mean dipole cosine & \(+0.38\) & \(-0.94\) & \(-0.82\) & \(-0.82\) \\
Negative dipole cosine & 5/18 & 17/17 & 16/17 & 16/17 \\
\bottomrule
\end{tabular}
\tabnotes{Pearson and Spearman coefficients are means of the within-molecule correlations against Hirshfeld for molecules with both charge-vector standard deviations above \(10^{-6}\,e\) (\(n=20\) for P; \(n=19\) for L, \(\mathrm{L}_{\mathrm{fit}}\), and Q, excluding SO\(_2\)). Balanced sign is \(\tfrac12(\mathrm{TPR}+\mathrm{TNR})\) on the 54 non-hydrogen atoms with \(|q^{\mathrm{Hirsh}}|\ge0.02\,e\). Dipole cosine is defined only when both norms exceed \(10^{-6}\,\mathrm{D}\) (\(n=18\) for P; \(n=17\) for L, \(\mathrm{L}_{\mathrm{fit}}\), and Q). Negative dipole cosine counts molecules with cosine \(<0\), i.e.\ an angle greater than \(90^\circ\). Two dipole errors are listed because they are not equivalent: the norm MAE \(\langle\,|\|\boldsymbol\mu_{\mathrm{EEM}}\|-\|\boldsymbol\mu_{\mathrm{DFT}}\|\,|\,\rangle\) is the primary score of Sec.~\ref{sec:verdicts}, while the vector MAE \(\langle\|\boldsymbol\mu_{\mathrm{EEM}}-\boldsymbol\mu_{\mathrm{DFT}}\|\rangle\) also penalizes direction. They rank the three geometric tables in opposite orders (L best on the norm, Q best on the vector), and the spread among the three, at most \(0.03\,\mathrm{D}\) either way, is an order of magnitude smaller than their common gap to P.}
\end{table}

\FloatBarrier

The charge error can fall while the dipole collapses, and the size of the predicted charges makes the two outcomes a single effect. Averaged over the same 117 atoms, \(\langle|q|\rangle\) is \(0.084\,e\) for Hirshfeld, \(0.099\,e\) for L, \(0.133\,e\) for Q, and \(0.200\,e\) for P. The Hirshfeld charges of this set are small, the flattened geometric contrasts produce charges of nearly the same size, and the spectroscopic contrasts of P overpolarize by a factor of about \(2.4\) against that reference. Ordering the tables by \(\langle|q|\rangle\) relative to Hirshfeld therefore reproduces the ordering of the charge MAE. The dipole weights the same charges by position, and the flattening that brings the L magnitudes into range is what removes the contrasts \(\boldsymbol{\mu}\) requires.

The spatial arrangement of charge within a molecule, as opposed to its magnitude, does not follow the error in the magnitude at all. Within each molecule of B2+B3 that has nonzero spread in both charge vectors, the Pearson correlation between the EEM and Hirshfeld charge vectors averages \(+0.107\) for P (\(n=20\); median \(+0.199\)) but \(-0.423\) for L (\(n=19\); median \(-0.637\)); Q behaves as L does, at \(-0.404\) (\(n=19\)). The complementary Spearman correlations are \(-0.021\) (P) and \(-0.560\) (L). SO\(_2\) is excluded under L, Q, and \(\mathrm{L}_{\mathrm{fit}}\) because those EEM charge vectors are identically zero. On the \(54\) non-hydrogen atoms carrying \(|q^{\mathrm{Hirsh}}|\ge0.02\,e\), P reproduces the Hirshfeld sign for \(50\%\) and L for \(37\%\). Because \(46\) of those \(54\) atoms are Hirshfeld-negative, an always-negative classifier would achieve \(46/54=85\%\) raw sign accuracy. To account for this class imbalance, the balanced sign rates \(\tfrac12(\mathrm{TPR}+\mathrm{TNR})\) are \(0.71\) for P and \(0.63\) for L. Figure~\ref{fig:place} makes the split visible: L is closer in absolute size and further in arrangement, and the dipole cosine follows the charge correlations rather than the charge MAE.


\FloatBarrier

\begin{figure}[!ht]
\centering
\includegraphics[width=0.92\textwidth]{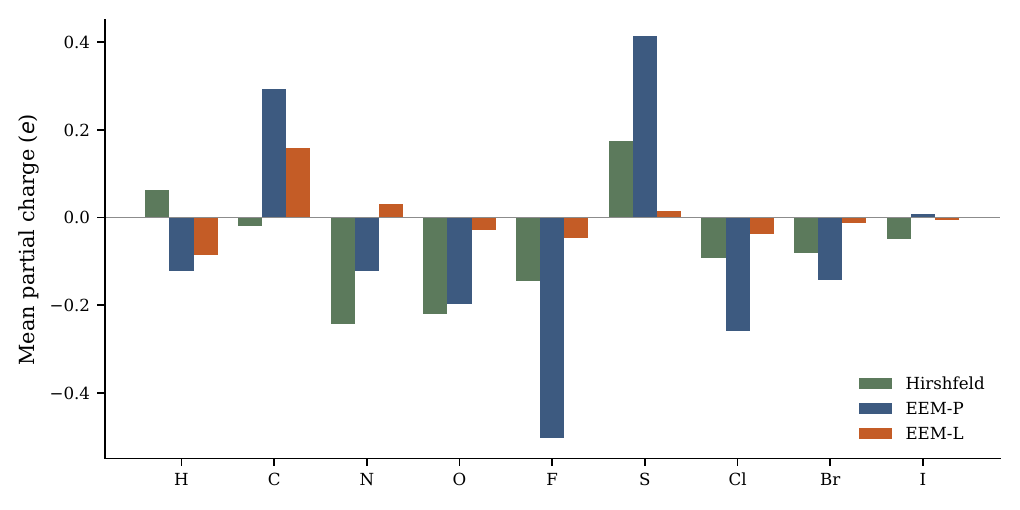}
\caption{Mean partial charge by element on B2+B3. Hirshfeld hydrogen is positive (\(+0.062\,e\)); L makes it negative (\(-0.086\,e\)), as does P (\(-0.123\,e\); means over the 60 hydrogens of B2+B3 from \texttt{data/verdicts.json}). Under the geometric table oxygen and nitrogen collapse towards neutrality, the molecular counterpart of the compressed \(\chi_L\) values in Fig.~\ref{fig:atomic}.}
\label{fig:element}
\end{figure}

\begin{figure}[!ht]
\centering
\includegraphics[width=0.92\textwidth]{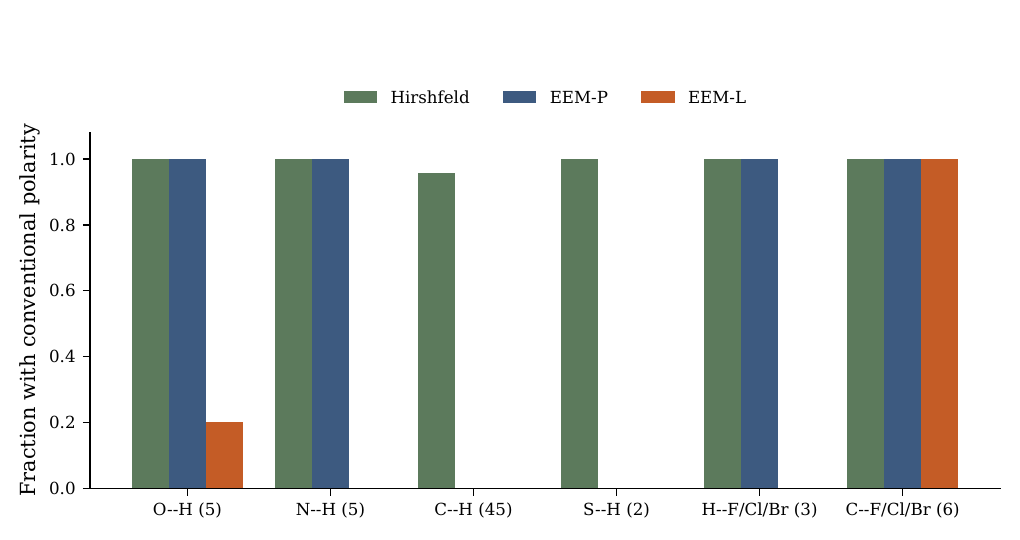}
\caption{Fraction of nearest-neighbor contacts on B1--B3 with conventional polarity (heavy atom more negative than hydrogen; carbon more positive than the halogen). Compression of the geometric table inverts the O--H, N--H and hydrogen-halide contacts, all of which the spectroscopic table and the Hirshfeld reference order conventionally. C--H is inverted under every table.}
\label{fig:bonds}
\end{figure}

Table~\ref{tab:atomic} and Fig.~\ref{fig:atomic} already place retained Pearson hydrogen above every geometric electronegativity. Table~P already places H above C, S, and P, so C--H polarity is inverted under every table; LiH in B1 and B--H are cases where hydrogen as the more electronegative partner is chemically conventional. The additional inversions created by L are those that P orders oppositely: O--H, N--H, and H--F. Figure~\ref{fig:element} shows the same inversion as an element-averaged trend: over the \(60\) hydrogens of B2+B3, \(q_{\mathrm{H}}\) is \(+0.062\,e\) for Hirshfeld and \(-0.086\,e\) for L, with carbon correspondingly positive (\(-0.019\,e\) for Hirshfeld, \(+0.158\,e\) for L). This element average is not on its own evidence against the geometric table, and it should be read with the spectroscopic column beside it: P gives \(-0.123\,e\) on the same hydrogens and \(+0.293\,e\) on the same carbons, so on this particular measure P lies further from the reference than L does. What separates the two tables is not the mean charge on hydrogen but which bonds carry the wrong sign, which is the comparison made below and in Fig.~\ref{fig:bonds}. Under the geometric table oxygen and nitrogen are left almost neutral, whereas both the Hirshfeld reference and the spectroscopic table keep them negative. Water, ammonia, and hydrogen peroxide have the DFT polarity under P and the reverse under L. The inversion survives the Ohno damping of Sec.~\ref{sec:eem} and is therefore a property of the table.


\FloatBarrier

Of the 117 atoms in B2+B3, 60 are hydrogen and 57 are geometric atoms. Atom-weighted MAE on hydrogen is \(0.184\,e\) (P) and \(0.148\,e\) (L); on the heavy atoms it is \(0.244\,e\) and \(0.179\,e\). Hydrogen is thus \(51\%\) of the atoms scored in the primary comparison, and it carries the same spectroscopic value in all four tables, so a reader is entitled to ask how much of the comparison is a test of the coordinate at all. Silver does not enter this set. The question is answerable: deleting hydrogen and rescoring the 57 remaining atoms gives atom-weighted errors of \(0.244\,e\) for P and \(0.179\,e\) for L, a ratio of \(0.73\) against \(0.76\) with hydrogen included. The shared hydrogen is therefore not what produces the smaller error of the geometric table, and removing it moves the ratio slightly further from unity rather than towards it. What the shared entries do limit is the interpretation: roughly half of the atom-level agreement in every table is inherited spectroscopic input common to all of them, so the charge comparison measures the coordinate only through the other half. Accordingly, the charge-magnitude comparison provides only limited evidence about the coordinate. The two hydrogen-free molecules, CO\(_2\) and SO\(_2\), reverse the ordering. Their two-stage MAE is \(0.076\,e\) for P and \(0.206\,e\) for L, because equal geometric electronegativities make the EEM charges on SO\(_2\) identically zero. The 18 hydrogen-containing molecules give \(0.242\,e\) (P) and \(0.172\,e\) (L). The polarity inversions should accordingly be attributed to the combination of the compressed periods-2--5 electronegativity scale with a hydrogen value imported unchanged from the spectroscopic table, rather than to the geometric values alone. Extending the construction to the first period would remove that asymmetry at its source.

Figure~\ref{fig:bonds} scores nearest-neighbor polarity on B1--B3 as the fraction of contacts in which the heavy atom is more negative than hydrogen (or, for C--X, carbon more positive than the halogen). Hirshfeld and P both put the heavy atom more negative than hydrogen in all five O--H contacts, all five N--H contacts, and in HF, HCl, and HBr. L does so in only one of the five O--H contacts and in none of the N--H or H--X contacts. C--H is inverted under every table, including P, because spectroscopic Mulliken already places H above C. S--H is likewise inverted under P as well as L. The three P--H contacts of phosphine are the one class in which the Hirshfeld reference is itself unconventional, placing phosphorus positive and its hydrogens negative; every table agrees with the reference there and none agrees with the textbook expectation, so the class separates nothing. C--F, C--Cl, and C--Br keep the conventional polarity in every table. C--I is mixed under Hirshfeld and P (\(0.5\)) and conventional under L. The polarity failures newly introduced by the compression of the geometric table are therefore O--H, N--H, and H--F/H--Cl/H--Br. The same hydrogen-sign pattern appears on the B8 alcohols, acids, and formamide (Supporting Information, the per-molecule charges quoted after Table~S5).


\FloatBarrier

Because the ordering by magnitude depends on the reference partitioning, the same two-stage error was recomputed against the Mulliken charges of the same wavefunctions. Those charges are larger, \(\langle|q|\rangle=0.177\,e\), and the errors rise to \(0.321\,e\) (P), \(0.279\,e\) (L), and \(0.306\,e\) (Q). L is still the closest of the three, but the ratio moves from \(0.779\) to \(0.868\), which lies inside the P/L charge criterion. The comparison of the geometric with the spectroscopic table on Hirshfeld magnitudes therefore depends on the partition chosen: the direction of the L/P comparison survives a change of partitioning while the operational rejection of the equivalence interval does not. A change of DFT recipe, as opposed to a change of partitioning, is considered in the Supporting Information (Table~S5). The sweep also tests sensitivity to molecule-set composition. It was run on a different panel of eight molecules, and on that panel the direction of the comparison reverses: \(\mathrm{MAE}(L)/\mathrm{MAE}(P)\) is \(1.048\), \(1.061\), \(1.036\) and \(1.053\) for the four recipes, so the geometric table is the worse of the two there under every recipe, and against Mulliken charges of the same densities the four ratios are \(1.22\), \(0.98\), \(1.23\) and \(1.35\). Read together with the bootstrap interval \([0.618,1.034]\), which contains unity, this places the charge advantage of the geometric table inside the scatter produced by exchanging a handful of molecules. The recipe is stable; the molecule list is not. The sweep does not recompute B2+B3 itself, so the robustness of the principal ratio is carried by the molecule-level bootstrap and the partition cross-check rather than by a recipe sweep of the same set.

\FloatBarrier

\subsection{Hard--soft fragment rankings}
\label{sec:e4}

The last comparison against reference data concerns the qualitative ordering of hard and soft species. Table~\ref{tab:hsab} ranks the four fragments by the response-mismatch rule of Eq.~\eqref{eq:mismatch} and asks whether each acid is placed nearer the base with which it is conventionally paired. The spectroscopic table places both acids nearer phosphine. The published geometric table places boron trifluoride nearer ammonia and borane nearer phosphine, which is the conventional pairing of hard with hard and soft with soft. The quadratic control and the refitted hyperbolic table both agree with the spectroscopic result instead. This is the one comparison in the paper that the published table passes and the spectroscopic table fails, and it is worth being precise about why. The success is a property of the published energy scales and not of the functional form, because the refitted table uses exactly the same hyperbolic kernels and does not reproduce the pairing. The mechanism is visible in the numbers. Under the spectroscopic table the two acids are almost degenerate, at \(7.95\) and \(8.00\,\mathrm{eV}\), and both lie below both bases, so the nearer base is phosphine in each row. Under the published table they separate by \(0.50\,\mathrm{eV}\), because borane contains only one atom whose hardness differs between the two tables whereas boron trifluoride contains four: hydrogen keeps its spectroscopic value, while the hardness of fluorine rises from \(7.01\) to \(7.90\,\mathrm{eV}\). Boron trifluoride is thereby lifted to \(8.87\,\mathrm{eV}\), within \(0.16\,\mathrm{eV}\) of ammonia, while borane reaches only \(8.37\,\mathrm{eV}\) and remains closest to phosphine. The mixed-basis finite-energy column places ammonia nearer both acids, because its four values lie within a narrow interval in which ammonia happens to be the closer base in each case. Importantly, the quantum-chemical reference agrees with the conventional assignment for boron trifluoride but disagrees with it for borane, so it too fails scoring rule~(3). A criterion that the reference calculation does not itself satisfy cannot carry much evidential weight, and we therefore do not read the single success of the published table as support for the coordinate. It is recorded because it was prespecified, not because it is decisive.

The extra AlCl\(_3\)/Ag\(^+\)/halide pairings are recorded in the Supporting Information; they do not form part of scoring rule~(3). Criterion~(3) is therefore treated as non-discriminating in Table~\ref{tab:verdicts} rather than as evidence for either atomic table.

\begin{table}[!ht]
\centering
\caption{Response-hardness ranking on the HSAB comparison.}
\label{tab:hsab}
\small
\begin{tabular}{lrrrrr}
\toprule
Fragment & \(K_{\mathrm{mol},P}\) & \(K_{\mathrm{mol},L}\) & \(K_{\mathrm{mol},\mathrm{Lfit}}\) & \(K_{\mathrm{mol},Q}\) & \(\eta_{\mathrm{MB}}\) \\
\midrule
BF\(_3\) & 7.95 & 8.87 & 8.26 & 8.26 & 7.90 \\
BH\(_3\) & 8.00 & 8.37 & 8.28 & 8.27 & 6.64 \\
NH\(_3\) & 8.99 & 9.03 & 9.10 & 9.10 & 6.17 \\
PH\(_3\) & 8.14 & 8.15 & 7.86 & 7.86 & 5.95 \\
\midrule
BF\(_3\) prefers & PH\(_3\) & NH\(_3\) & PH\(_3\) & PH\(_3\) & NH\(_3\) \\
BH\(_3\) prefers & PH\(_3\) & PH\(_3\) & PH\(_3\) & PH\(_3\) & NH\(_3\) \\
\bottomrule
\end{tabular}
\tabnotes{Values are in eV: EEM response slopes \(\Kmol\) follow Eq.~\eqref{eq:etamol}; the mixed-basis finite-energy column follows Eqs.~\eqref{eq:mb} and~\eqref{eq:mixed-basis-ea}. Preference is defined by Eq.~\eqref{eq:mismatch} among NH\(_3\) and PH\(_3\). \(\mathrm{L}_{\mathrm{fit}}\) agrees with P and Q, not with published-scale L. The first four columns and the last are not on the same scale and must not be compared across the table. \(\Kmol\) is twice a hardness, since Eq.~\eqref{eq:etamol} returns \(2\eta\) for an isolated atom, whereas \(\eta_{\mathrm{MB}}\) is a hardness. The like-for-like quantity is \(\eta_{\mathrm{mol}}=\Kmol/2\), which for these four fragments is \(3.98\), \(4.00\), \(4.50\) and \(4.07\,\mathrm{eV}\) under the spectroscopic table against \(7.90\), \(6.64\), \(6.17\) and \(5.95\,\mathrm{eV}\) for \(\eta_{\mathrm{MB}}\). Each column is used only for the ordering within itself, which is what Eq.~\eqref{eq:mismatch} requires.}
\end{table}

\subsection{Molecular hardness from atomic values}
\label{sec:e5}
The final comparison is a negative control rather than a test of any particular table: it asks how well molecular hardness can be assembled from atomic values at all, and the answer turns out to be much the same for all four tables. The mixed-basis reference inherits the one-sided basis bias described in Sec.~\ref{sec:dft}, so no scoring verdict rests on this subsection.

Table~\ref{tab:hard} compares combination rules with finite \(\eta_{\mathrm{MB}}\) values for 47 species. Per-species indices are in Table~S4. Excluded are N\(_2\) and CH\(_4\), whose \((N+1)\) jobs were omitted because those anions are known unbound, the halide ions F\(^-\) and I\(^-\), for which adding a further electron would produce a dianion, and KCl, which has a negative EEM response slope under L, Q, and \(\mathrm{L}_{\mathrm{fit}}\). The cause is the size of the hardness mismatch in that molecule: potassium is assigned \(0.532\,\mathrm{eV}\) by the geometric table and \(0.475\,\mathrm{eV}\) by the quadratic control against \(5.579\,\mathrm{eV}\) for chlorine, a factor of ten, which is exactly the circumstance that Sec.~\ref{sec:etamol} identifies as able to drive \(\Kmol\) negative while the charge solution itself remains a minimum. The negative response in KCl is not an isolated molecular anomaly. The hardness of the geometric table is proportional to the inward gap \(\kappaRS\), which is largest for the atom next to the closing noble gas and smallest at the opposite edge of the period. The alkali metals are that opposite edge by definition: lithium and sodium sit at \(\rho=7/8\) and potassium at \(\rho=17/18\), the largest values in their periods, and they receive the smallest hardnesses in the table, \(1.909\), \(1.347\) and \(0.532\,\mathrm{eV}\) against \(2.387\), \(2.296\) and \(1.920\,\mathrm{eV}\) in the spectroscopic table. Every atom opposite a closing noble gas is compressed in the same way, so any alkali halide will show a large hardness mismatch and the effect grows down the group. Potassium chloride is where the mismatch first becomes large enough to change the sign of the response slope; it is the visible instance of a systematic feature of the coordinate, not an isolated artifact. The vertical NO\(^{-}\) calculation uses the triplet ground state and is included, as are the two B7 diatomics. Negative electron affinities among the remaining species are retained as finite-basis values. The arithmetic mean of atomic \(\eta_L\) has MAE \(0.838\,\mathrm{eV}\), essentially the same as the Pearson atomic mean (\(0.822\,\mathrm{eV}\)). The like-for-like EEM comparison is \(\eta_{\mathrm{mol}}=\Kmol/2\) versus \(\eta_{\mathrm{MB}}\): the MAEs are \(2.32\), \(2.24\), and \(2.39\,\mathrm{eV}\) for P, L, and Q, against an \(\eta_{\mathrm{MB}}\) range of \(4.0\)--\(8.9\,\mathrm{eV}\) across these species, so the error is about half the full spread of the quantity being predicted. EEM molecular hardness performs poorly against this mixed-basis finite-energy index for every table, not only L. That is a negative result for this Ohno-EEM response index measured against this particular reference, and it does not generalize: against the frontier-orbital index \(\eta_{\mathrm{HL}}\) the same \(\eta_{\mathrm{mol}}\) has MAE \(0.66\), \(0.80\) and \(0.79\,\mathrm{eV}\) for P, L and Q, the smallest errors in Table~\ref{tab:hard} and roughly a third of those of the atomic arithmetic means against the same reference. The two references therefore disagree about the response index on identical predictions and identical species, so at most one of them is measuring the quantity \(\Kmol\) represents. We do not claim it is the frontier-orbital one: \(\eta_{\mathrm{HL}}\) rests on Koopmans-type orbital energies whose absolute scale is not an observable, so the close agreement is a consistency check and not a validation. What is robust to the choice of hardness reference is that P, L and Q remain within \(0.14\,\mathrm{eV}\) of one another in the new column; this comparison therefore does not distinguish the tables. Raw \(\Kmol\) versus \(\eta_{\mathrm{MB}}\) happens to give smaller numbers (\(1.43\), \(1.91\), and \(1.71\,\mathrm{eV}\)), but those quantities use different conventions. Off-diagonal coupling has already pulled \(\Kmol\) down to a mean of \(7.3\,\mathrm{eV}\) under P (\(7.5\) under L, \(7.1\) under Q) against \(6.0\,\mathrm{eV}\) for \(\eta_{\mathrm{MB}}\), about \(1.2\) times rather than twice, so the two indices are separated by the coupling and not by a missing factor of two. Against \(\eta_{\mathrm{HL}}\), atomic-mean \(\eta_L\) has MAE \(2.50\,\mathrm{eV}\). The remaining combination rule promised in Sec.~\ref{sec:etamol}, Sanderson's geometric mean of the atomic \(\chi_L\) values against \(\chi_{\mathrm{MB}}\), has MAE \(1.71\,\mathrm{eV}\) on the same species. Only the L value was recorded, so this rule is reported for the geometric table alone and is not one of the table-to-table comparisons.

\begin{table}[!ht]
\centering
\caption{Combination-rule MAE in eV for the 47 species scored in Sec.~\ref{sec:e5}.}
\label{tab:hard}
\small
\begin{tabular}{lrr}
\toprule
Predictor & vs \(\eta_{\mathrm{MB}}\) & vs \(\eta_{\mathrm{HL}}\) \\
\midrule
Arithmetic mean \(\eta_P\) & 0.822 & 2.038 \\
Arithmetic mean \(\eta_L\) & 0.838 & 2.496 \\
Arithmetic mean \(\eta_Q\) & 0.832 & 2.101 \\
EEM \(\eta_{\mathrm{mol}}=\Kmol/2\) (P) & 2.321 & 0.663 \\
EEM \(\eta_{\mathrm{mol}}=\Kmol/2\) (L) & 2.242 & 0.802 \\
EEM \(\eta_{\mathrm{mol}}=\Kmol/2\) (Q) & 2.391 & 0.791 \\
EEM response \(\Kmol\) (P) & 1.434 & 3.681 \\
EEM response \(\Kmol\) (L) & 1.908 & 3.908 \\
EEM response \(\Kmol\) (Q) & 1.712 & 3.657 \\
\bottomrule
\end{tabular}
\tabnotes{Of the 52 species in sets B1--B7, five are excluded: N\(_2\) and CH\(_4\) (no \((N+1)\) job), F\(^-\) and I\(^-\), for which electron addition would give a dianion, and KCl (negative compressed-table EEM response slope). The 47 therefore already exclude KCl. The vertical NO\(^{-}\) state is the triplet. The principal EEM comparison is the like-for-like \(\eta_{\mathrm{mol}}=\Kmol/2\). Raw \(\Kmol\) versus \(\eta_{\mathrm{MB}}\) uses a different normalization. Negative electron affinities are retained. The \(\eta_{\mathrm{HL}}\) entries for \(\eta_{\mathrm{mol}}\) are computed for the same 47 species from \texttt{data/hardness.csv} as the other entries.}
\end{table}

\FloatBarrier

\subsection{Is the degeneracy removable?}
\label{sec:repair}
A further question follows from the failures reported above: whether the framework is repairable at all or is unsound in a way no amount of additional information would fix. One diagnostic settles it, reported here without offering the result as a table for use. The degeneracy has a single identifiable cause. Electronegativity in the published construction is one global constant times a function of the coordinate, \(\chi_L=C_\chi\,\chistruct\), so two atoms with the same coordinate in different periods must receive the same value; hardness escapes this because its constant \(C_\eta(p)\) is already resolved by period. The asymmetry between the two scales is thus the whole source of this degeneracy, and the minimal repair is to remove it by fitting \(C_\chi(p)\) period by period, through the origin, by exactly the protocol already used for the hardness. Doing so gives \(129.84\), \(103.07\), \(130.49\) and \(113.77\,\mathrm{eV}\) for periods~2 to~5, and it separates all seven degenerate pairs by \(0.75\) to \(1.63\,\mathrm{eV}\), fluorine from chlorine by \(1.63\,\mathrm{eV}\) and oxygen from sulfur by \(1.60\,\mathrm{eV}\). At the molecular level, the period-resolved scaling has the following consequences. For the five polyatomic heavy-atom--hydrogen tests, the polarity inversion disappears completely: the heavy-atom-to-hydrogen polarity of water, ammonia, methanol, methylamine and hydrogen peroxide is reproduced in five cases out of five, against one out of five for the published table and five out of five for the spectroscopic one, and the hydrogen halides improve from none of four to two of four. The charge error on the twenty polyatomic molecules falls from \(0.175\,e\) to \(0.113\,e\), better than any of the four tables scored above, though still just short of the \(0.106\,e\) of the zero-charge model. Two conclusions follow. The shared-coordinate electronegativity degeneracy is not intrinsic to locating atoms by their distance to the closing noble gas; it follows from using one electronegativity scale for the whole table, and the coordinate survives the repair. But the repair is not a vindication either: it adds one fitted constant per period, taking the count from one to four, it still does not beat a model that assigns no charge at all, it leaves hydrogen outside the construction, and its period-5 constant is fixed by iodine alone and so is not an independent datum, exactly as the hardness constant is not. The same limitation also affects period~4, where the constant is fitted from potassium and bromine alone. The fifteen atoms carrying a coordinate are distributed \(6\), \(6\), \(2\) and \(1\) across periods~2 to~5, so the four constants of Eq.~\eqref{eq:Lscales} are not comparably determined: the first two are each fitted to six atoms, whereas \(C_\chi(4)\) rests on potassium and bromine alone and \(C_\chi(5)\) on iodine alone. Two of the four constants are therefore fixed by three atoms between them. This shows in the values, since \(C_\chi(4)=130.49\,\mathrm{eV}\) is the largest of the four and exceeds the period-2 value of \(129.84\,\mathrm{eV}\), breaking the otherwise monotonic decrease down the table; on two atoms that non-monotonicity cannot be distinguished from noise. The repair should accordingly be read as well determined for periods~2 and~3 and provisional below them. We report the diagnostic because a framework that fails for a locatable and removable reason is a different object from one that fails irreparably, and the distinction matters to anyone deciding whether to develop the idea further. The constants, the pairwise splittings and the recomputed charges are in the Supporting Information.

As an additional molecular check, the repaired table was also applied to set B8, which played no part in fitting the period-resolved constants: those are set on atomic data alone. Its two-stage charge error there is \(0.095\,e\), against \(0.182\,e\) for the published geometric table, \(0.246\,e\) for the spectroscopic table and \(0.104\,e\) for the zero-charge model. The repaired table also places a positive charge on the hydrogen at all five hydroxyl and amide contacts of that set, in agreement with the sign of the Hirshfeld reference, whereas the published table makes every one of them negative. Eight molecules are a first check rather than a validation, and the repaired charges remain compressed in magnitude.

\section{Discussion}
\label{sec:discussion}

Table~\ref{tab:verdicts} summarizes the six criteria of Sec.~\ref{sec:verdicts} so that their joint interpretation can be checked against the stated scoring rules.

\begin{table}[!ht]
\centering
\caption{Outcome of the six criteria of Sec.~\ref{sec:verdicts}.}
\label{tab:verdicts}
\small
\begin{tabular}{@{}lccccc@{\hspace{0.9em}}l@{}}
\toprule
Criterion & P & L & \(\mathrm{L}_{\mathrm{fit}}\) & Q & Reference & Separates L from controls? \\
\midrule
(1) charges closer than the zero-charge model & no & no & no & no & --- & no, all four fail \\
(2) HX dipole order and sign & no & no & no & no & yes & no, all four fail \\
(3) both conventional pairings (post hoc form) & no & yes & no & no & no & yes, but reference fails \\
\midrule
\multicolumn{7}{@{}p{0.97\textwidth}@{}}{(A) \(\mathrm{MAE}(L)/\mathrm{MAE}(P)\) in \([0.8,1.2]\): unresolved: the ratio is \(0.779\), but its interval \([0.62,1.03]\) covers both the equivalence-interval edge and unity, and against Mulliken charges the ratio is \(0.868\), inside the interval} \\
\multicolumn{7}{@{}p{0.97\textwidth}@{}}{(B) matched kernels agree to \(5\%\): met, at \(0.0011\) on charges and \(0.0089\) on dipoles; published-scale L against Q gives \(0.114\), and that published-scale form is the prespecified one} \\
\multicolumn{7}{@{}p{0.97\textwidth}@{}}{(C) nonvanishing dipole for ClF and IBr: failed by L, \(\mathrm{L}_{\mathrm{fit}}\) and Q, each giving exactly zero; P gives \(1.79\) and \(1.53\,\mathrm{D}\)} \\
\bottomrule
\end{tabular}
\tabnotes{The upper block compares each table with reference data; the Reference column gives the corresponding reference-calculation outcome where that question has an answer. The final column records whether the criterion separates L from the controls. The lower block compares tables with one another and has a single outcome apiece. The criteria are abbreviated here and stated in full in Sec.~\ref{sec:verdicts}: (1) is the two-stage charge error on B2+B3 against the model that puts zero charge on every atom, (2) is the descent of the HX dipole norms from F to I with the sign given by density-functional theory, and (3) is the pairing of BF\(_3\) with NH\(_3\) and BH\(_3\) with PH\(_3\). None of the three does so in a way that bears on the coordinate. Criteria~(1) and~(2) are failed by all four tables alike; criterion~(1) remains informative as a benchmark against the null model, but it does not distinguish L from the other frozen tables. Criterion~(3) is met by published-scale L alone, but it is failed by the reference calculation as well, so meeting it carries no evidential weight, and the conventional-pairing form shown here was introduced after the outcome was known and is a post hoc diagnostic; the prespecified form asked whether L reproduces the P ranking (Sec.~\ref{sec:e4}). The rows should be weighted accordingly rather than counted as three failures. Criterion~(A) is recorded as unresolved rather than rejected because its bootstrap interval covers the equivalence-interval edge (Sec.~\ref{sec:e1}). Criterion~(B) is reported in both forms, and the prespecified form is the one using published-scale L, which fails; the matched \(\mathrm{L}_{\mathrm{fit}}\) form was added after that outcome was known and is a diagnostic, not a criterion.}
\end{table}

\FloatBarrier

The matched-kernel comparison has implications beyond the specific noble-gas table. The geometric construction is motivated in part by the functional form used to map the structural coordinate to atomic parameters. Here that mapping is hyperbolic with base \(\phig\). When that form and an ordinary quadratic are given the same freedom, fitted to the same spectroscopic values by the same through-origin protocol, and then applied to the same twenty molecules, they agree to \(0.0011\) on charges and \(0.0089\) on dipole moments in units of the spectroscopic error. The tolerance set in advance was \(5\%\), so the charge agreement is about one fiftieth of it and the dipole agreement about one sixth; both are far below the scatter between molecule lists. Within the coordinate range, the fitting protocol and the equalization construction examined here, the molecular predictions are therefore controlled primarily by the fitted energy scales rather than by the detailed shape of the kernel. Mathematical distinctness of a kernel therefore does not, by itself, establish molecular relevance without an independently discriminating molecular test.

Accordingly, the golden-ratio kernel receives no support from these molecular data, consistent with its lack of identification from the free-atom values. The companion mathematical results establish structural properties of the kernel, whereas the present calculation asks whether that mathematical distinctness produces discriminating molecular predictions. On the molecule sets examined here it does not: mathematical uniqueness of a functional form and empirical identification of that form are separate claims.

The second negative finding is algebraic rather than numerical, and it is worth setting out plainly. The atomic degeneracy it starts from was documented in the paper that proposed the coordinate; what is added here is that the degeneracy reaches a molecular observable, and reaches it exactly, with no fitted quantity, no reference calculation and no choice of partition entering the argument. For a fixed hardness matrix the equalization equations are linear in the atomic electronegativities, so if all of those electronegativities are equal in a neutral molecule then zero charge on every atom satisfies the equations together with the charge constraint, as shown in Sec.~\ref{sec:eem}. The collapse of the dipole moment for molecules built entirely from one degeneracy class is thus a property of the table and not an accident of the numerical solver, and it cannot be repaired by adjusting the energy scales or the damping. Chlorine monofluoride, iodine monobromide and sulfur dioxide are the instances that contain no hydrogen, and for all three the measured and computed dipole moments are finite while the geometric table predicts exactly zero (Fig.~\ref{fig:carry}). It should be noted that changing the table also changes the hardness matrix, so the comparisons reported here are always the effect of a whole table and not of its electronegativities in isolation. It is also worth distinguishing what is being tested here from what is normally done: the usual practice is to fit effective electronegativities and hardnesses to quantum-chemical charges, whereas we freeze a table constructed in advance. The negative result is that the frozen table does not behave like those fitted effective parameters.

Three of the results carry an interpretation that goes beyond their own numbers. First, both failure modes belong to the atomic table rather than to the solver, and neither is removed by changing the damping of Eq.~\eqref{eq:ohno}: the collapse of the dipole moment is the algebraic consequence set out above, while the inversion of O--H, N--H and hydrogen-halide polarity arises from retaining spectroscopic hydrogen above the compressed heavy-atom values (Figs.~\ref{fig:element} and~\ref{fig:bonds}). Period-resolved scaling mitigates this problem, as shown in Sec.~\ref{sec:repair}, while extending the construction to the first period would remove the asymmetry at its source. Second, the comparisons that the geometric table wins are the ones with the least diagnostic value. Its smaller charge error on the twenty polyatomic molecules depends on the choice of partition and is in any case beaten by placing zero charge on every atom (Sec.~\ref{sec:e1}, Tables~\ref{tab:charge} and~\ref{tab:place}), while the conventional hard--soft pairing that it does reproduce is not reproduced by the same kernel with refitted scales, so that success belongs to the published energy scales. Third, the poor performance of the molecular response index against the finite-energy reference is shared by all four tables (Sec.~\ref{sec:e5}) and is therefore a property of that comparison rather than of the coordinate, carrying no information about the coordinate. Taken together, the comparisons that discriminate between tables reject the geometric table, and those in which it appears competitive do not discriminate. That summary is worth making specific, because only one of the six criteria both separates the tables and goes against the geometric one, and it is the vanishing dipole moment of criterion~(C). Criterion~(1) separates every table from the zero-charge model rather than from another table; criterion~(2) is failed by all four and separates nothing; criterion~(3) favours the geometric table but is failed by the reference calculation; criterion~(A) is left unresolved by its own interval; and criterion~(B) is a statement about kernels and not about the coordinate. The negative verdict of this paper therefore rests on criterion~(C) together with the position of all four tables relative to the zero-charge model, and not on an accumulation of failures.

The comparison also separates limitations of frozen atomic tables in general from failures specific to the noble-gas construction. The spectroscopic table P is itself frozen and likewise lies above the zero-charge Hirshfeld baseline on B2+B3, but it shows neither the exact shared-coordinate dipole collapse nor the additional O--H and N--H polarity inversions. The shortfall in charge magnitude is therefore not peculiar to the noble-gas coordinate in this equalization form, whereas the structural degeneracy is. Fitted and environment-dependent equalization models lie outside the present test.

 The design limitations of the study are collected in Sec.~\ref{sec:scope}; two of them bear directly on the findings above. The first is the composition of the molecule list. The additional set of eight organic molecules shows that the equivalence of the two kernels, the ordering relative to the zero-charge model and the inversion of hydrogen polarity are not peculiar to the original list. Two eight-molecule panels appear in this paper and should not be conflated: set B8 contains no degenerate pair, so it cannot test the collapse, whereas the recipe sweep panel of Table~S5 contains sulfur dioxide, so the central phenomenon of this paper is present there and survives every change of functional, basis set and partition reference. On the eight molecules of that sweep panel the charge-error ratio rises above unity under every recipe, and we therefore regard the smaller charge error of the geometric table on B2+B3 as a property of that particular list rather than as a transferable result (Sec.~\ref{sec:e1}). The second is the reference itself. Varying the functional, the basis set or the reference densities of the partition moves the reference charges by at most \(0.0042\,e\), so the recipe is not the source of the effects reported here; the choice of partition, rather than the recipe, remains the principal reference-related caveat on the comparison of charge magnitudes.

\section{Conclusions}
\label{sec:conclusions}

Inserted unchanged into the fixed-geometry equalization model used here, the noble-gas table fails as a frozen atomic input. Its two principal molecular failures can be traced to features of the table construction rather than to the equalization solver. Because the coordinate is built from the distance to the closing noble gas and the length of the period and from nothing else, any two atoms with the same ratio of the two receive one and the same electronegativity. In the table used here that happens for the seven pairs enumerated in Sec.~\ref{sec:rho}, of which the present molecule list can probe three. The hardnesses within a pair are not degenerate, since the hardness scale is resolved by period, but it is the electronegativities alone that drive the equalization equations, so any neutral molecule assembled entirely from a single degeneracy class has a dipole moment of exactly zero. Because the first period lies outside the construction, hydrogen keeps a spectroscopic electronegativity above every compressed value, which inverts the polarity of O--H, N--H and hydrogen-halide bonds. A separate hardness-scale diagnostic shows that the near-vanishing hardness assigned at the far edge of a period can also produce an unphysical response, KCl giving a negative molecular response slope under all three compressed tables. Neither of the two principal features is a matter of calibration: after identical fitting the hyperbolic and quadratic forms agree atom by atom to \(0.0022\,e\), which is \(0.98\%\) of the spectroscopic charge error (Sec.~\ref{sec:e2q}), and on the polyatomic molecules of sets B2 and B3 no frozen table comes as close to the reference charge magnitudes as assigning zero charge to every atom, although on the ionic species of sets B1 and B6 every table beats that model. We do not infer that L is more or less accurate than P for the charge-magnitude comparison. On the twenty polyatomic molecules the error of L is the smaller of the two, but the interval on the paired difference contains zero and the interval on the ratio contains both the equivalence boundary and unity, the ordering reverses against a second density partition, and it reverses again on a further eight-molecule panel. None of the findings below depends on it. The case rests on rule~(1) and on the vanishing dipole moment, for the reasons given in Sec.~\ref{sec:discussion}. A generally usable replacement would require information beyond the two integers that define the coordinate, sufficient to separate fluorine from chlorine, oxygen from sulfur and bromine from iodine, together with a rule that places hydrogen on the same footing as the atoms the construction already describes.

The degeneracy is removable in principle, and Sec.~\ref{sec:repair} reports the diagnostic that shows it: resolving the electronegativity scale period by period, as the hardness scale already is, separates all seven degenerate pairs, restores the correct polarity of all five polyatomic heavy-atom--hydrogen tests, and lowers the charge error to \(0.113\,e\). The cost is three further fitted constants. The repaired table remains above the zero-charge model on B2+B3, but falls below it on the eight molecules of set B8, where it also restores the reference sign of the hydrogen charge at all five hydroxyl and amide contacts. The degeneracy identified here is therefore repairable, but the additional fitted constants and the eight-molecule B8 check do not establish a generally transferable replacement table.

\begin{suppinfo}
The following files are available free of charge.
\begin{itemize}
\item \texttt{Molecular\_Descriptors\_EEM\_SI.pdf}: per-molecule organohalide dipoles, charge MAEs, dipole norms, and hardness indices (Tables~S1--S4); the frozen-geometry sweep over functional, basis set and partition reference (Table~S5); the four frozen atomic tables at the precision actually used, so that every equalization result can be regenerated from the published inputs (Table~S6); the paired comparison of each frozen table with the zero-charge model on the twenty polyatomic molecules, with Wilcoxon tests and bootstrap intervals (Table~S7); the period-resolved electronegativity diagnostic, with the fitted constants, the resulting separation of all seven degenerate pairs and the additional check on set B8 (Table~S8); and the two-stage charge error as the base of the hyperbolic kernel is swept continuously (Table~S9); together with the per-molecule B8 charges reported in the text and notes on bootstrap, Hessian, and dipole-convention checks, including the smallest constrained-Hessian eigenvalue found under each table.
\item \texttt{Molecular\_Descriptors\_EEM\_data\_and\_code.zip}: optimized Cartesian coordinates, atomic \((\chi,\eta)\) tables, scoring scripts, and the numerical summary that regenerates the main-text tables.
\end{itemize}
\end{suppinfo}

\begin{acknowledgement}
Electronic-structure calculations used {PySCF}~2.7.0 and {geomeTRIC}~1.1. This work received no external funding.
\end{acknowledgement}

\bibliography{Molecular_Descriptors_EEM}

\providecommand{\latin}[1]{#1}
\makeatletter
\providecommand{\doi}
  {\begingroup\let\do\@makeother\dospecials
  \catcode`\{=1 \catcode`\}=2 \doi@aux}
\providecommand{\doi@aux}[1]{\endgroup\texttt{#1}}
\makeatother
\providecommand*\mcitethebibliography{\thebibliography}
\csname @ifundefined\endcsname{endmcitethebibliography}
  {\let\endmcitethebibliography\endthebibliography}{}
\begin{mcitethebibliography}{52}
\providecommand*\natexlab[1]{#1}
\providecommand*\mciteSetBstSublistMode[1]{}
\providecommand*\mciteSetBstMaxWidthForm[2]{}
\providecommand*\mciteBstWouldAddEndPuncttrue
  {\def\EndOfBibitem{\unskip.}}
\providecommand*\mciteBstWouldAddEndPunctfalse
  {\let\EndOfBibitem\relax}
\providecommand*\mciteSetBstMidEndSepPunct[3]{}
\providecommand*\mciteSetBstSublistLabelBeginEnd[3]{}
\providecommand*\EndOfBibitem{}
\mciteSetBstSublistMode{f}
\mciteSetBstMaxWidthForm{subitem}{(\alph{mcitesubitemcount})}
\mciteSetBstSublistLabelBeginEnd
  {\mcitemaxwidthsubitemform\space}
  {\relax}
  {\relax}

\bibitem[Rapp{\'e} and Goddard~III(1991)Rapp{\'e}, and
  Goddard~III]{rappe1991qeq}
Rapp{\'e},~A.~K.; Goddard~III,~W.~A. Charge equilibration for molecular
  dynamics simulations. \emph{J. Phys. Chem.} \textbf{1991}, \emph{95},
  3358--3363\relax
\mciteBstWouldAddEndPuncttrue
\mciteSetBstMidEndSepPunct{\mcitedefaultmidpunct}
{\mcitedefaultendpunct}{\mcitedefaultseppunct}\relax
\EndOfBibitem
\bibitem[Verstraelen \latin{et~al.}(2009)Verstraelen, Van~Speybroeck, and
  Waroquier]{verstraelen2009sqe}
Verstraelen,~T.; Van~Speybroeck,~V.; Waroquier,~M. The electronegativity
  equalization method and the split charge equilibration applied to organic
  systems: parametrization, validation, and comparison. \emph{J. Chem. Phys.}
  \textbf{2009}, \emph{131}, 044127, DOI: \doi{10.1063/1.3187034}\relax
\mciteBstWouldAddEndPuncttrue
\mciteSetBstMidEndSepPunct{\mcitedefaultmidpunct}
{\mcitedefaultendpunct}{\mcitedefaultseppunct}\relax
\EndOfBibitem
\bibitem[Verstraelen \latin{et~al.}(2013)Verstraelen, Ayers, Van~Speybroeck,
  and Waroquier]{verstraelen2013acks2}
Verstraelen,~T.; Ayers,~P.~W.; Van~Speybroeck,~V.; Waroquier,~M. {ACKS2}:
  Atom-condensed {K}ohn--{S}ham {DFT} approximated to second order. \emph{J.
  Chem. Phys.} \textbf{2013}, \emph{138}, 074108, DOI:
  \doi{10.1063/1.4791569}\relax
\mciteBstWouldAddEndPuncttrue
\mciteSetBstMidEndSepPunct{\mcitedefaultmidpunct}
{\mcitedefaultendpunct}{\mcitedefaultseppunct}\relax
\EndOfBibitem
\bibitem[Pauling(1932)]{pauling1932nature}
Pauling,~L. The nature of the chemical bond. {IV}. The energy of single bonds
  and the relative electronegativity of atoms. \emph{J. Am. Chem. Soc.}
  \textbf{1932}, \emph{54}, 3570--3582\relax
\mciteBstWouldAddEndPuncttrue
\mciteSetBstMidEndSepPunct{\mcitedefaultmidpunct}
{\mcitedefaultendpunct}{\mcitedefaultseppunct}\relax
\EndOfBibitem
\bibitem[Allred and Rochow(1958)Allred, and
  Rochow]{allred1958electronegativity}
Allred,~A.~L.; Rochow,~E.~G. A scale of electronegativity based on
  electrostatic force. \emph{J. Inorg. Nucl. Chem.} \textbf{1958}, \emph{5},
  264--268\relax
\mciteBstWouldAddEndPuncttrue
\mciteSetBstMidEndSepPunct{\mcitedefaultmidpunct}
{\mcitedefaultendpunct}{\mcitedefaultseppunct}\relax
\EndOfBibitem
\bibitem[Allen(1989)]{allen1989electronegativity}
Allen,~L.~C. Electronegativity is the average one-electron energy of the
  valence-shell electrons in ground-state free atoms. \emph{J. Am. Chem. Soc.}
  \textbf{1989}, \emph{111}, 9003--9014\relax
\mciteBstWouldAddEndPuncttrue
\mciteSetBstMidEndSepPunct{\mcitedefaultmidpunct}
{\mcitedefaultendpunct}{\mcitedefaultseppunct}\relax
\EndOfBibitem
\bibitem[Sanderson(1951)]{sanderson1951electronegativity}
Sanderson,~R.~T. An interpretation of bond lengths and a classification of
  bonds. \emph{Science} \textbf{1951}, \emph{114}, 670--672\relax
\mciteBstWouldAddEndPuncttrue
\mciteSetBstMidEndSepPunct{\mcitedefaultmidpunct}
{\mcitedefaultendpunct}{\mcitedefaultseppunct}\relax
\EndOfBibitem
\bibitem[Sanderson(1983)]{sanderson1983electronegativity}
Sanderson,~R.~T. Electronegativity and bond energy. \emph{J. Am. Chem. Soc.}
  \textbf{1983}, \emph{105}, 2259--2261\relax
\mciteBstWouldAddEndPuncttrue
\mciteSetBstMidEndSepPunct{\mcitedefaultmidpunct}
{\mcitedefaultendpunct}{\mcitedefaultseppunct}\relax
\EndOfBibitem
\bibitem[Mortier \latin{et~al.}(1985)Mortier, Van~Genechten, and
  Gasteiger]{mortier1985electronegativity}
Mortier,~W.~J.; Van~Genechten,~K.; Gasteiger,~J. Electronegativity
  equalization: application and parametrization. \emph{J. Am. Chem. Soc.}
  \textbf{1985}, \emph{107}, 829--835\relax
\mciteBstWouldAddEndPuncttrue
\mciteSetBstMidEndSepPunct{\mcitedefaultmidpunct}
{\mcitedefaultendpunct}{\mcitedefaultseppunct}\relax
\EndOfBibitem
\bibitem[Mortier \latin{et~al.}(1986)Mortier, Ghosh, and
  Shankar]{mortier1986electronegativity}
Mortier,~W.~J.; Ghosh,~S.~K.; Shankar,~S. Electronegativity equalization method
  for the calculation of atomic charges in molecules. \emph{J. Am. Chem. Soc.}
  \textbf{1986}, \emph{108}, 4315--4320\relax
\mciteBstWouldAddEndPuncttrue
\mciteSetBstMidEndSepPunct{\mcitedefaultmidpunct}
{\mcitedefaultendpunct}{\mcitedefaultseppunct}\relax
\EndOfBibitem
\bibitem[Bultinck \latin{et~al.}(2002)Bultinck, Langenaeker, Lahorte, De~Proft,
  Geerlings, Waroquier, and Tollenaere]{bultinck2002eem}
Bultinck,~P.; Langenaeker,~W.; Lahorte,~P.; De~Proft,~F.; Geerlings,~P.;
  Waroquier,~M.; Tollenaere,~J.~P. The electronegativity equalization method
  {I}: parametrization and validation for atomic charge calculations. \emph{J.
  Phys. Chem. A} \textbf{2002}, \emph{106}, 7887--7894\relax
\mciteBstWouldAddEndPuncttrue
\mciteSetBstMidEndSepPunct{\mcitedefaultmidpunct}
{\mcitedefaultendpunct}{\mcitedefaultseppunct}\relax
\EndOfBibitem
\bibitem[Nistor \latin{et~al.}(2006)Nistor, Polihronov, M{\"u}ser, and
  Mosey]{nistor2006split}
Nistor,~R.~A.; Polihronov,~J.~G.; M{\"u}ser,~M.~H.; Mosey,~N.~J. A
  generalization of the charge equilibration method for nonmetallic materials.
  \emph{J. Chem. Phys.} \textbf{2006}, \emph{125}, 094108, DOI:
  \doi{10.1063/1.2346671}\relax
\mciteBstWouldAddEndPuncttrue
\mciteSetBstMidEndSepPunct{\mcitedefaultmidpunct}
{\mcitedefaultendpunct}{\mcitedefaultseppunct}\relax
\EndOfBibitem
\bibitem[Mathieu(2007)]{mathieu2007split}
Mathieu,~D. Split charge equilibration method with correct dissociation limits.
  \emph{J. Chem. Phys.} \textbf{2007}, \emph{127}, 224103, DOI:
  \doi{10.1063/1.2803060}\relax
\mciteBstWouldAddEndPuncttrue
\mciteSetBstMidEndSepPunct{\mcitedefaultmidpunct}
{\mcitedefaultendpunct}{\mcitedefaultseppunct}\relax
\EndOfBibitem
\bibitem[Nalewajski \latin{et~al.}(1996)Nalewajski, Korchowiec, and
  Michalak]{nalewajski1996charge}
Nalewajski,~R.~F.; Korchowiec,~J.; Michalak,~A. Reactivity criteria in charge
  sensitivity analysis. \emph{Top. Curr. Chem.} \textbf{1996}, \emph{183},
  25--141\relax
\mciteBstWouldAddEndPuncttrue
\mciteSetBstMidEndSepPunct{\mcitedefaultmidpunct}
{\mcitedefaultendpunct}{\mcitedefaultseppunct}\relax
\EndOfBibitem
\bibitem[Verstraelen \latin{et~al.}(2011)Verstraelen, Bultinck, Van~Speybroeck,
  Ayers, Van~Neck, and Waroquier]{verstraelen2011params}
Verstraelen,~T.; Bultinck,~P.; Van~Speybroeck,~V.; Ayers,~P.~W.; Van~Neck,~D.;
  Waroquier,~M. The significance of parameters in charge equilibration models.
  \emph{J. Chem. Theory Comput.} \textbf{2011}, \emph{7}, 1750--1764, DOI:
  \doi{10.1021/ct200006e}\relax
\mciteBstWouldAddEndPuncttrue
\mciteSetBstMidEndSepPunct{\mcitedefaultmidpunct}
{\mcitedefaultendpunct}{\mcitedefaultseppunct}\relax
\EndOfBibitem
\bibitem[Ohno(1964)]{ohno1964}
Ohno,~K. Some remarks on the {P}ariser--{P}arr--{P}ople method. \emph{Theor.
  Chim. Acta} \textbf{1964}, \emph{2}, 219--227\relax
\mciteBstWouldAddEndPuncttrue
\mciteSetBstMidEndSepPunct{\mcitedefaultmidpunct}
{\mcitedefaultendpunct}{\mcitedefaultseppunct}\relax
\EndOfBibitem
\bibitem[Washburn \latin{et~al.}(2026)Washburn, Simons, and
  Allahyarov]{washburn2026noblegas}
Washburn,~J.; Simons,~M.; Allahyarov,~E. A noble gas-centered coordinate for
  within-period atomic property trends. \emph{Symmetry} \textbf{2026},
  \emph{18}, 1087, DOI: \doi{10.3390/sym18071087}\relax
\mciteBstWouldAddEndPuncttrue
\mciteSetBstMidEndSepPunct{\mcitedefaultmidpunct}
{\mcitedefaultendpunct}{\mcitedefaultseppunct}\relax
\EndOfBibitem
\bibitem[Ko \latin{et~al.}(2021)Ko, Finkler, Goedecker, and
  Behler]{ko2021fourth}
Ko,~T.~W.; Finkler,~J.~A.; Goedecker,~S.; Behler,~J. A fourth-generation
  high-dimensional neural network potential with accurate electrostatics
  including non-local charge transfer. \emph{Nat. Commun.} \textbf{2021},
  \emph{12}, 398, DOI: \doi{10.1038/s41467-020-20427-2}\relax
\mciteBstWouldAddEndPuncttrue
\mciteSetBstMidEndSepPunct{\mcitedefaultmidpunct}
{\mcitedefaultendpunct}{\mcitedefaultseppunct}\relax
\EndOfBibitem
\bibitem[Mulliken(1934)]{mulliken1934electronegativity}
Mulliken,~R.~S. A new electroaffinity scale; together with data on valence
  states and on valence ionization potentials and electron affinities. \emph{J.
  Chem. Phys.} \textbf{1934}, \emph{2}, 782--793\relax
\mciteBstWouldAddEndPuncttrue
\mciteSetBstMidEndSepPunct{\mcitedefaultmidpunct}
{\mcitedefaultendpunct}{\mcitedefaultseppunct}\relax
\EndOfBibitem
\bibitem[Pearson(1988)]{pearson1988absolute}
Pearson,~R.~G. Absolute electronegativity and hardness: application to
  inorganic chemistry. \emph{Inorg. Chem.} \textbf{1988}, \emph{27},
  734--740\relax
\mciteBstWouldAddEndPuncttrue
\mciteSetBstMidEndSepPunct{\mcitedefaultmidpunct}
{\mcitedefaultendpunct}{\mcitedefaultseppunct}\relax
\EndOfBibitem
\bibitem[Iczkowski and Margrave(1961)Iczkowski, and
  Margrave]{iczkowski1961electronegativity}
Iczkowski,~R.~P.; Margrave,~J.~L. Electronegativity. \emph{J. Am. Chem. Soc.}
  \textbf{1961}, \emph{83}, 3547--3551\relax
\mciteBstWouldAddEndPuncttrue
\mciteSetBstMidEndSepPunct{\mcitedefaultmidpunct}
{\mcitedefaultendpunct}{\mcitedefaultseppunct}\relax
\EndOfBibitem
\bibitem[Parr \latin{et~al.}(1978)Parr, Donnelly, Levy, and
  Palke]{parr1978electronegativity}
Parr,~R.~G.; Donnelly,~R.~A.; Levy,~M.; Palke,~W.~E. Electronegativity: the
  density functional viewpoint. \emph{J. Chem. Phys.} \textbf{1978}, \emph{68},
  3801--3807\relax
\mciteBstWouldAddEndPuncttrue
\mciteSetBstMidEndSepPunct{\mcitedefaultmidpunct}
{\mcitedefaultendpunct}{\mcitedefaultseppunct}\relax
\EndOfBibitem
\bibitem[Parr and Pearson(1983)Parr, and Pearson]{parr1983absolute}
Parr,~R.~G.; Pearson,~R.~G. Absolute hardness: companion parameter to absolute
  electronegativity. \emph{J. Am. Chem. Soc.} \textbf{1983}, \emph{105},
  7512--7516\relax
\mciteBstWouldAddEndPuncttrue
\mciteSetBstMidEndSepPunct{\mcitedefaultmidpunct}
{\mcitedefaultendpunct}{\mcitedefaultseppunct}\relax
\EndOfBibitem
\bibitem[Pearson(1963)]{pearson1963hard}
Pearson,~R.~G. Hard and soft acids and bases. \emph{J. Am. Chem. Soc.}
  \textbf{1963}, \emph{85}, 3533--3539\relax
\mciteBstWouldAddEndPuncttrue
\mciteSetBstMidEndSepPunct{\mcitedefaultmidpunct}
{\mcitedefaultendpunct}{\mcitedefaultseppunct}\relax
\EndOfBibitem
\bibitem[Pearson(1968)]{pearson1968hsab}
Pearson,~R.~G. Hard and soft acids and bases, {HSAB}, part {I}: fundamental
  principles. \emph{J. Chem. Educ.} \textbf{1968}, \emph{45}, 581--587\relax
\mciteBstWouldAddEndPuncttrue
\mciteSetBstMidEndSepPunct{\mcitedefaultmidpunct}
{\mcitedefaultendpunct}{\mcitedefaultseppunct}\relax
\EndOfBibitem
\bibitem[Chattaraj \latin{et~al.}(1991)Chattaraj, Lee, and
  Parr]{chattaraj1991hsab}
Chattaraj,~P.~K.; Lee,~H.; Parr,~R.~G. {HSAB} principle. \emph{J. Am. Chem.
  Soc.} \textbf{1991}, \emph{113}, 1855--1856\relax
\mciteBstWouldAddEndPuncttrue
\mciteSetBstMidEndSepPunct{\mcitedefaultmidpunct}
{\mcitedefaultendpunct}{\mcitedefaultseppunct}\relax
\EndOfBibitem
\bibitem[Ayers \latin{et~al.}(2006)Ayers, Parr, and Pearson]{ayers2006hsab}
Ayers,~P.~W.; Parr,~R.~G.; Pearson,~R.~G. Elucidating the hard/soft acid/base
  principle: a perspective based on half-reactions. \emph{J. Chem. Phys.}
  \textbf{2006}, \emph{124}, 194107, DOI: \doi{10.1063/1.2196882}\relax
\mciteBstWouldAddEndPuncttrue
\mciteSetBstMidEndSepPunct{\mcitedefaultmidpunct}
{\mcitedefaultendpunct}{\mcitedefaultseppunct}\relax
\EndOfBibitem
\bibitem[Washburn and Zlatanovi{\'c}(2026)Washburn, and
  Zlatanovi{\'c}]{washburn2026uniqueness}
Washburn,~J.; Zlatanovi{\'c},~M. Uniqueness of the canonical reciprocal cost.
  \emph{Mathematics} \textbf{2026}, \emph{14}, 935, DOI:
  \doi{10.3390/math14060935}\relax
\mciteBstWouldAddEndPuncttrue
\mciteSetBstMidEndSepPunct{\mcitedefaultmidpunct}
{\mcitedefaultendpunct}{\mcitedefaultseppunct}\relax
\EndOfBibitem
\bibitem[Washburn and Rahnamai~Barghi(2026)Washburn, and
  Rahnamai~Barghi]{washburn2026reciprocal}
Washburn,~J.; Rahnamai~Barghi,~A. Reciprocal convex costs for ratio matching:
  axiomatic characterization. \emph{Axioms} \textbf{2026}, \emph{15}, 151, DOI:
  \doi{10.3390/axioms15020151}\relax
\mciteBstWouldAddEndPuncttrue
\mciteSetBstMidEndSepPunct{\mcitedefaultmidpunct}
{\mcitedefaultendpunct}{\mcitedefaultseppunct}\relax
\EndOfBibitem
\bibitem[Kramida \latin{et~al.}(2024)Kramida, Ralchenko, Reader, and {NIST ASD
  Team}]{nist_asd}
Kramida,~A.; Ralchenko,~Y.; Reader,~J.; {NIST ASD Team}, {NIST} Atomic Spectra
  Database, version 5.12. National Institute of Standards and Technology,
  Gaithersburg, MD, 2024; Accessed August 13, 2026\relax
\mciteBstWouldAddEndPuncttrue
\mciteSetBstMidEndSepPunct{\mcitedefaultmidpunct}
{\mcitedefaultendpunct}{\mcitedefaultseppunct}\relax
\EndOfBibitem
\bibitem[Hotop and Lineberger(1985)Hotop, and Lineberger]{hotop1985electron}
Hotop,~H.; Lineberger,~W.~C. Binding energies in atomic negative ions: {II}.
  \emph{J. Phys. Chem. Ref. Data} \textbf{1985}, \emph{14}, 731--750\relax
\mciteBstWouldAddEndPuncttrue
\mciteSetBstMidEndSepPunct{\mcitedefaultmidpunct}
{\mcitedefaultendpunct}{\mcitedefaultseppunct}\relax
\EndOfBibitem
\bibitem[Andersen \latin{et~al.}(1999)Andersen, Haugen, and
  Hotop]{andersen1999binding}
Andersen,~T.; Haugen,~H.~K.; Hotop,~H. Binding energies in atomic negative
  ions: {III}. \emph{J. Phys. Chem. Ref. Data} \textbf{1999}, \emph{28},
  1511--1533\relax
\mciteBstWouldAddEndPuncttrue
\mciteSetBstMidEndSepPunct{\mcitedefaultmidpunct}
{\mcitedefaultendpunct}{\mcitedefaultseppunct}\relax
\EndOfBibitem
\bibitem[Ning and Lu(2022)Ning, and Lu]{ning2022ea}
Ning,~C.; Lu,~Y. Electron affinities of atoms and structures of atomic negative
  ions. \emph{J. Phys. Chem. Ref. Data} \textbf{2022}, \emph{51}, 021502, DOI:
  \doi{10.1063/5.0080243}\relax
\mciteBstWouldAddEndPuncttrue
\mciteSetBstMidEndSepPunct{\mcitedefaultmidpunct}
{\mcitedefaultendpunct}{\mcitedefaultseppunct}\relax
\EndOfBibitem
\bibitem[Nalewajski \latin{et~al.}(1988)Nalewajski, Korchowiec, and
  Zhou]{nalewajski1988hardness}
Nalewajski,~R.~F.; Korchowiec,~J.; Zhou,~Z. Molecular hardness and softness
  parameters and their use in chemistry. \emph{Int. J. Quantum Chem.}
  \textbf{1988}, \emph{34}, 349--366, DOI: \doi{10.1002/qua.560340840}\relax
\mciteBstWouldAddEndPuncttrue
\mciteSetBstMidEndSepPunct{\mcitedefaultmidpunct}
{\mcitedefaultendpunct}{\mcitedefaultseppunct}\relax
\EndOfBibitem
\bibitem[Fukui \latin{et~al.}(1952)Fukui, Yonezawa, and
  Shingu]{fukui1952reactivity}
Fukui,~K.; Yonezawa,~T.; Shingu,~H. A molecular orbital theory of reactivity in
  aromatic hydrocarbons. \emph{J. Chem. Phys.} \textbf{1952}, \emph{20},
  722--725\relax
\mciteBstWouldAddEndPuncttrue
\mciteSetBstMidEndSepPunct{\mcitedefaultmidpunct}
{\mcitedefaultendpunct}{\mcitedefaultseppunct}\relax
\EndOfBibitem
\bibitem[Parr and Yang(1984)Parr, and Yang]{parr1984fukui}
Parr,~R.~G.; Yang,~W. Density functional approach to the frontier-electron
  theory of chemical reactivity. \emph{J. Am. Chem. Soc.} \textbf{1984},
  \emph{106}, 4049--4050\relax
\mciteBstWouldAddEndPuncttrue
\mciteSetBstMidEndSepPunct{\mcitedefaultmidpunct}
{\mcitedefaultendpunct}{\mcitedefaultseppunct}\relax
\EndOfBibitem
\bibitem[Yang and Mortier(1986)Yang, and Mortier]{yang1986fukui}
Yang,~W.; Mortier,~W.~J. The use of global and local molecular parameters for
  the analysis of the gas-phase basicity of amines. \emph{J. Am. Chem. Soc.}
  \textbf{1986}, \emph{108}, 5708--5711\relax
\mciteBstWouldAddEndPuncttrue
\mciteSetBstMidEndSepPunct{\mcitedefaultmidpunct}
{\mcitedefaultendpunct}{\mcitedefaultseppunct}\relax
\EndOfBibitem
\bibitem[Fuentealba \latin{et~al.}(2000)Fuentealba, P{\'e}rez, and
  Contreras]{fuentealba2000fukui}
Fuentealba,~P.; P{\'e}rez,~P.; Contreras,~R. On the condensed {F}ukui function.
  \emph{J. Chem. Phys.} \textbf{2000}, \emph{113}, 2544--2551\relax
\mciteBstWouldAddEndPuncttrue
\mciteSetBstMidEndSepPunct{\mcitedefaultmidpunct}
{\mcitedefaultendpunct}{\mcitedefaultseppunct}\relax
\EndOfBibitem
\bibitem[Yang \latin{et~al.}(1985)Yang, Lee, and Ghosh]{yang1985softness}
Yang,~W.; Lee,~C.; Ghosh,~S.~K. Molecular softness as the average of atomic
  softnesses: companion principle to the geometric mean principle for
  electronegativity equalization. \emph{J. Phys. Chem.} \textbf{1985},
  \emph{89}, 5412--5414\relax
\mciteBstWouldAddEndPuncttrue
\mciteSetBstMidEndSepPunct{\mcitedefaultmidpunct}
{\mcitedefaultendpunct}{\mcitedefaultseppunct}\relax
\EndOfBibitem
\bibitem[Sun \latin{et~al.}(2018)Sun, \latin{et~al.} others]{sun2018pyscf}
Sun,~Q., \latin{et~al.}  {PySCF}: the {P}ython-based simulations of chemistry
  framework. \emph{WIREs Comput. Mol. Sci.} \textbf{2018}, \emph{8},
  e1340\relax
\mciteBstWouldAddEndPuncttrue
\mciteSetBstMidEndSepPunct{\mcitedefaultmidpunct}
{\mcitedefaultendpunct}{\mcitedefaultseppunct}\relax
\EndOfBibitem
\bibitem[Sun \latin{et~al.}(2020)Sun, \latin{et~al.} others]{sun2020pyscf}
Sun,~Q., \latin{et~al.}  Recent developments in the {PySCF} program package.
  \emph{J. Chem. Phys.} \textbf{2020}, \emph{153}, 024109, DOI:
  \doi{10.1063/5.0006074}\relax
\mciteBstWouldAddEndPuncttrue
\mciteSetBstMidEndSepPunct{\mcitedefaultmidpunct}
{\mcitedefaultendpunct}{\mcitedefaultseppunct}\relax
\EndOfBibitem
\bibitem[Wang and Song(2016)Wang, and Song]{wang2016geometry}
Wang,~L.-P.; Song,~C. Geometry optimization made simple with translation and
  rotation coordinates. \emph{J. Chem. Phys.} \textbf{2016}, \emph{144},
  214108\relax
\mciteBstWouldAddEndPuncttrue
\mciteSetBstMidEndSepPunct{\mcitedefaultmidpunct}
{\mcitedefaultendpunct}{\mcitedefaultseppunct}\relax
\EndOfBibitem
\bibitem[Becke(1993)]{becke1993density}
Becke,~A.~D. Density-functional thermochemistry. {III}. The role of exact
  exchange. \emph{J. Chem. Phys.} \textbf{1993}, \emph{98}, 5648--5652\relax
\mciteBstWouldAddEndPuncttrue
\mciteSetBstMidEndSepPunct{\mcitedefaultmidpunct}
{\mcitedefaultendpunct}{\mcitedefaultseppunct}\relax
\EndOfBibitem
\bibitem[Lee \latin{et~al.}(1988)Lee, Yang, and Parr]{lee1988lyp}
Lee,~C.; Yang,~W.; Parr,~R.~G. Development of the {C}olle--{S}alvetti
  correlation-energy formula into a functional of the electron density.
  \emph{Phys. Rev. B} \textbf{1988}, \emph{37}, 785--789, DOI:
  \doi{10.1103/PhysRevB.37.785}\relax
\mciteBstWouldAddEndPuncttrue
\mciteSetBstMidEndSepPunct{\mcitedefaultmidpunct}
{\mcitedefaultendpunct}{\mcitedefaultseppunct}\relax
\EndOfBibitem
\bibitem[Vosko \latin{et~al.}(1980)Vosko, Wilk, and Nusair]{vosko1980vwn}
Vosko,~S.~H.; Wilk,~L.; Nusair,~M. Accurate spin-dependent electron liquid
  correlation energies for local spin density calculations: a critical
  analysis. \emph{Can. J. Phys.} \textbf{1980}, \emph{58}, 1200--1211, DOI:
  \doi{10.1139/p80-159}\relax
\mciteBstWouldAddEndPuncttrue
\mciteSetBstMidEndSepPunct{\mcitedefaultmidpunct}
{\mcitedefaultendpunct}{\mcitedefaultseppunct}\relax
\EndOfBibitem
\bibitem[Stephens \latin{et~al.}(1994)Stephens, Devlin, Chabalowski, and
  Frisch]{stephens1994}
Stephens,~P.~J.; Devlin,~F.~J.; Chabalowski,~C.~F.; Frisch,~M.~J. Ab initio
  calculation of vibrational absorption and circular dichroism spectra using
  density functional force fields. \emph{J. Phys. Chem.} \textbf{1994},
  \emph{98}, 11623--11627\relax
\mciteBstWouldAddEndPuncttrue
\mciteSetBstMidEndSepPunct{\mcitedefaultmidpunct}
{\mcitedefaultendpunct}{\mcitedefaultseppunct}\relax
\EndOfBibitem
\bibitem[Weigend and Ahlrichs(2005)Weigend, and Ahlrichs]{weigend2005}
Weigend,~F.; Ahlrichs,~R. Balanced basis sets of split valence, triple zeta
  valence and quadruple zeta valence quality for {H} to {R}n. \emph{Phys. Chem.
  Chem. Phys.} \textbf{2005}, \emph{7}, 3297--3305\relax
\mciteBstWouldAddEndPuncttrue
\mciteSetBstMidEndSepPunct{\mcitedefaultmidpunct}
{\mcitedefaultendpunct}{\mcitedefaultseppunct}\relax
\EndOfBibitem
\bibitem[Rappoport and Furche(2010)Rappoport, and Furche]{rappoport2010def2d}
Rappoport,~D.; Furche,~F. Property-optimized {G}aussian basis sets for
  molecular response calculations. \emph{J. Chem. Phys.} \textbf{2010},
  \emph{133}, 134105, DOI: \doi{10.1063/1.3484283}\relax
\mciteBstWouldAddEndPuncttrue
\mciteSetBstMidEndSepPunct{\mcitedefaultmidpunct}
{\mcitedefaultendpunct}{\mcitedefaultseppunct}\relax
\EndOfBibitem
\bibitem[Johnson~III(2022)]{cccbdb}
Johnson~III,~R.~D. {NIST} Computational Chemistry Comparison and Benchmark
  Database, {NIST} Standard Reference Database 101, release 22. 2022;
  \url{https://cccbdb.nist.gov/}, accessed August 14, 2026\relax
\mciteBstWouldAddEndPuncttrue
\mciteSetBstMidEndSepPunct{\mcitedefaultmidpunct}
{\mcitedefaultendpunct}{\mcitedefaultseppunct}\relax
\EndOfBibitem
\bibitem[Bak \latin{et~al.}(2000)Bak, Gauss, Helgaker, J{\o}rgensen, and
  Olsen]{bak2000dipole}
Bak,~K.~L.; Gauss,~J.; Helgaker,~T.; J{\o}rgensen,~P.; Olsen,~J. The accuracy
  of molecular dipole moments in standard electronic structure calculations.
  \emph{Chem. Phys. Lett.} \textbf{2000}, \emph{319}, 563--568\relax
\mciteBstWouldAddEndPuncttrue
\mciteSetBstMidEndSepPunct{\mcitedefaultmidpunct}
{\mcitedefaultendpunct}{\mcitedefaultseppunct}\relax
\EndOfBibitem
\bibitem[Hirshfeld(1977)]{hirshfeld1977}
Hirshfeld,~F.~L. Bonded-atom fragments for describing molecular charge
  densities. \emph{Theor. Chim. Acta} \textbf{1977}, \emph{44}, 129--138\relax
\mciteBstWouldAddEndPuncttrue
\mciteSetBstMidEndSepPunct{\mcitedefaultmidpunct}
{\mcitedefaultendpunct}{\mcitedefaultseppunct}\relax
\EndOfBibitem
\end{mcitethebibliography}

\end{document}